\documentclass[aps, prb,twocolumn, superscriptaddress,amsmath, amssymb, longbibliography]{revtex4-2} 
\usepackage{graphicx}
\usepackage{amsmath}
\usepackage{dcolumn}
\usepackage{bm}
\usepackage{lmodern}
\usepackage[T1]{fontenc}
\usepackage{esint}

\usepackage[colorlinks,bookmarks=false,citecolor=blue,linkcolor=blue,urlcolor=blue]{hyperref}
\usepackage{xurl}
\hypersetup{breaklinks=true}

\begin{document}

\title{Collisionless dynamics of superconducting gap excited by spin-splitting field}

\author{V.~Plastovets}
\affiliation{University of Bordeaux, LOMA UMR-CNRS 5798, F-33405 Talence Cedex, France}
\affiliation{Institute for Physics of Microstructures, Russian Academy of Sciences, 603950 Nizhny Novgorod, GSP-105, Russia}
\author{A.~S.~Mel'nikov}
\affiliation{Institute for Physics of Microstructures, Russian Academy of Sciences, 603950 Nizhny Novgorod, GSP-105, Russia}
\affiliation{Moscow Institute of Physics and Technology, 141701 Dolgoprudny, Russia}
\affiliation{Lobachevsky State University of Nizhny Novgorod, 603950 Nizhni Novgorod, Russia}
\author{A.~I.~Buzdin}
\affiliation{University of Bordeaux, LOMA UMR-CNRS 5798, F-33405 Talence Cedex, France}

\date{\today}

\begin{abstract}
We study the coherent dynamic interaction of a time-dependent spin-splitting field with the homogeneous superconducting order parameter $\Delta(t)$ mediated by spin-orbit coupling  using the time-dependent Bogoliubov-de Gennes theory. \textit{In the first part} of the work we show that linear response of the superconductor is strongly affected by the Zeeman field and spin-flip processes, giving rise to multiple resonant frequencies of the superconducting Higgs modes. These modes can be excited either by a quench, or by an additional non-stationary component of the spin-splitting field, which couples linearly to the Higgs modes. \textit{In the second part}, we analyze the nonadiabatic dynamics of quasiparticle states arising from the intersection of spectral branches from different spin subbands, which can be provoked by a linearly growing Zeeman field. We provide insights into the dependence of the order parameter $\Delta(t)$ on this field and interference effects caused by tunneling of states at the avoided crossing points. We also show that since the nonadiabatic tunneling is related to spin-flip processes, the quasiparticle gas experiences a dynamic magnetization that contributes to its spin susceptibility. 
\end{abstract}

\maketitle

\section{Introduction}

Extensive studies of nonequilibrium states of superconductors \cite{langenberg1986nonequilibrium, kopnin2001theory} pay considerable attention to the so-called collisionless dynamics of a superconducting condensate, described by the complex-valued pairing potential $\Delta(t)$. At timescales shorter than the typical inelastic relaxation time $t\ll \tau_\varepsilon$ the dynamics of Cooper pairs is in coherent regime and is described by the Keldysh technique for Green's functions or its quasiclassical approximation \cite{1974JETP38.1018V,kulik_pair_1981}. The collisionless regime manifests itself most clearly in the existence of oscillations of the amplitude of the order parameter $\Delta(t)=\Delta_0+\delta\Delta(t)$ near the equilibrium gap value $\Delta_0$. This mode comes from excited interference interaction between the wavefunctions of the quasiparticles (QP) from broken Cooper pairs.Due to the QP dispersion the summation over all interference contributions results in an inhomogeneous broadening of the total gap mode, which is equivalent to a weak damping with a typical time evolution $\delta \Delta(t)\propto \cos(2\Delta_0 t)/\sqrt{\Delta_0t}$ \cite{1974JETP38.1018V}. By analogy with electro-weak particle theory this amplitude mode is called Higgs mode \cite{PhysRevLett.13.508, pekker_amplitudehiggs_2015}. Since the Higgs mode is a scalar excitation, it can not be coupled to the electromagnetic field ${\bf A}(t)$ linearly and several indirect mechanisms have been studied, such as linear excitation by the THz radiation in the presence of dc supercurrent \cite{PhysRevLett.118.047001, PhysRevLett.122.257001}. Also it possible to realize a nonlinear and coherent (or incoherent \cite{PhysRevB.105.104513}) Higgs mode excitation using high-intensity THz light with frequency just above the equilibrium superconducting gap $\Delta_0$, which can be detected by ultrafast pump-probe spectroscopy and third harmonic generation measurements \cite{PhysRevB.76.224522,doi:10.1126/science.1254697,PhysRevLett.111.057002, PhysRevB.92.224517, doi:10.1146/annurev-conmatphys-031119-050813}. 

It is known that, in addition to electromagnetic fields, superconductors also respond to nonstationary spin-splitting fields ${\bf h}(t)$. Typically, this field is produced by an external magnetic field ${\bf h}=\mu_B {\bf H}$ or by the exchange field of an adjacent ferromagnetic layer ${\bf h}\sim J_{\text{ex}}{\bf m}$, which is induced by proximity to the superconductor. Spin-split systems serve as a good platforms for spintronic applications and extensive study of various non-equilibrium processes has been done over the last few decades \cite{buzdin_proximity_2005, eschrig_spin-polarized_2015, heikkila_thermal_2019}. In particular, by inducing magnetic moment dynamics in S/F junctions an effective spin-triplet component of the superconducting gap is generated resulting in long-range proximity effects \cite{houzet_ferromagnetic_2008, Barnes_2011, PhysRevB.80.220502, PhysRevLett.99.057003}. On the other hand experimental observations indicate that the superconducting subsystem has a direct impact on the ferromagnetic resonance in hybrid S/F structures \cite{Li_2018, PhysRevApplied.19.034025, PhysRevApplied.18.L061004}.

In recent years there has been a growing interest in studying of the Higgs modes in the proximized superconducting systems \cite{vadimov_higgs_2019, PhysRevResearch.2.022068} as well as the interaction of collective modes in S/F systems. For instance, it was recently shown that in a superconductor in the helical phase, which can be achieved in the presence of a strong spin-orbit coupling (SOC) and an exchange field, the Higgs mode can be linearly coupled to the electromagnetic field through the nonzero superconducting phase gradient in the ground state \cite{https://doi.org/10.48550/arxiv.2212.11615}. Also, it was revealed that the coupling of the Higgs mode $\delta\Delta(t)$ in a superconductor to external light ${\bf A}(t)$ and magnetic dynamics ${\bf m}(t)$ in the F layer allows the generation of time-dependent spin currents \cite{silaev_spin_2020}. These currents can themselves excite the Higgs mode in the superconductor through the resonance of the ferromagnet due to the reciprocal effect \cite{silaev_spin_2020}. Another example is an interplay between the superconducting Higgs mode and a magnon mode in the adjacent F layer in the presence of a SOC and static proximity effect \cite{lu_coupling_2022}. Interestingly, the Higgs mode here is coupled to the Zeeman field ${\bf h}(t)$ linearly due to the presence of both the spin-orbit interaction and some preferred direction given by wave-vector of the magnetic mode. 

According to the aforementioned works, the SOC is critical for interaction of different spin subbands of the QP spectrum, which directly leads to the gap dynamics $\Delta(t)$. Some preconditions for this can be taken from the elementary analysis of the equilibrium state. The equilibrium superconducting gap does not depend on the Zeeman field below the so-called paramagnetic limit $h_\text{cr}=\Delta_0/\sqrt 2$, so that $\Delta(h<h_\text{cr}, T=0)=\Delta_0$; and above this limit  the  superconductivity is completely suppressed with $\Delta(h>h_\text{cr}, T=0)=0$ \cite{SARMA19631029, AbrikosovBook}. The SOC drastically changes the dependence $\Delta(h)$ and promotes a generation of triplet component superconducting correlations, leading to the survival of the gap at $h>h_\text{cr}$ \cite{Tewari_2011}. This effect is associated with mixing of the different spin states of QP, and the appearance of such mixing is naturally expected in the dynamic regime. 

The general description of dynamics of the superconducting condensate in the presence of both time-dependent spin-flipping field and SOC is rather difficult problem \cite{silaev_spin_2020,lu_coupling_2022}. In this paper we will focus on a fairly simple and specific system configuration, which allows us to explicitly trace the temporal evolution of QP states and its contribution to the order parameter $\Delta(t)$. For the sake of simplicity we consider an uniform superconductor at zero temperature $T=0$ and consider short timescale $t\ll\tau_\varepsilon$ at which the collisionless regime holds, so one can treat the system with the pure quantum-mechanical approach within the time-dependent Bogoliubov-de Gennes (TDBdG) equations \cite{KettersonBook}. We confine ourselves to addressing the homogeneous spin-splitting field with only one component ${\bf h}(t)=h(t){\bf z}_0$. A simple approach based on the expansion of the QP wave function in terms of the eigenstates of the BdG Hamiltonian $\psi(t)=\sum_n C_{n}(t) \Psi_{n}$ can be developed. The behavior of the QPs and related self-consistent gap function $\Delta(t)$ are determined by the coefficients $C_{n}(t)$, which describe how the states with a specific spin quantum number and momentum are refilled due to nonstationary transitions. After introducing the TDBdG equations in  Sec. \ref{sec_2},  we examine two different regimes of coherent evolution of the order parameter. 

In Sec. \ref{sec_3} we analyze linearized gap dynamics where the temporal evolution of the Higgs modes $\delta\Delta(t)$ is traced in the presence of a stationary spin-splitting field $h_0$. Since the SOC allows the transitions between the QP states with different spins, the induced perturbation of the gap $\delta\Delta(t)$ acquires three eigenfrequencies including the standard $2\Delta_0$ and two additional frequencies $2(\Delta_0\pm h_0)$. These modes define both the free oscillation of the perturbed gap at $h_0^{-1}\ll t \ll \tau_\varepsilon$, and resonant peaks in the case of driven oscillations. It was also shown that in the specific configuration, the linear coupling of the Higgs mode and the perturbation of the Zeeman field $\delta{\bf h}(t)$ co-directional with ${\bf h}_0$ is possible. Note that the frequencies shifted by the spin-splitting field have been observed in the numerical simulation of the dynamics of the one-dimensional Fermi superfluid exposed to the nonstationary Zeeman field and strong SOC in the Ref. \cite{Fan:52502}.

In Sec. \ref{sec_4} we consider the dynamics of the gap $\Delta(t)$ driven by linearly growing field ${\bf h}(t)$. At some point the field becomes larger then the equilibrium gap value $|{\bf h}(t)|>\Delta_0$ and thus provokes the crossing of the branches from different spin subbands of the QP spectrum. The appearance of non-adiabatic transitions between the states at the intersection point is equivalent to the dynamical spin-flip process and can be described with the Landau-Zener-St\"{u}ckelberg-Majorana (LZSM) tunneling problem \cite{ivakhnenko_nonadiabatic_2023}. Corresponding refilling of the amplitudes $C_n(t)$ contributes to the gap function $\Delta(t)$ and can drastically change its behavior depending on the field growth rate. In Sec. \ref{sec_4_C}-\ref{sec_4_D} we derive an analytical expression for $\Delta(t)$ from the self-consistency equation which contains two different terms: (i) smooth dependence $\Delta_h[h(t)]$ arising directly from spin-flip tunneling and depending on the occupation probability of QP states $|C_n(t)|$; (ii) oscillating part $\delta\Delta(t)$ originated from the interference between redistributed states with terms of type $C_n^*(t)C_{n'}(t)$. In addition, in Sec. \ref{sec_4_E}-\ref{sec_4_F} we discussed the spin imbalance generated due to LZSM tunneling and corresponding dynamical magnetization of the QP gas. Discussion and some experimental proposals are presented in Sec. \ref{sec_5}.

\section{Time-dependent Bogoliubov -- de Gennes equations }\label{sec_2}

We consider a homogeneous s-wave superconductor in the presence of the uniform time-dependent Zeeman field $h(t)$ and Rashba spin-orbit coupling (RSOC). 
The coherent QP dynamics is governed by the TDBdG equations \cite{KettersonBook}
\begin{gather}\label{Hamiltonian_BdG_SOI}
i\frac{\partial}{\partial t} \check{\psi}_k
= \check{\mathcal{H}}(k, t) \check{\psi}_k,
\end{gather}
where the Hamiltonian
\begin{gather}\label{TDHam}
\check{\mathcal{H}}(k, t) =
\begin{pmatrix}
\hat{H}(k,t)   &    i\hat{\sigma}_y\Delta(t)    \\
-i\hat{\sigma}_y\Delta(t)  &     -\hat{H}^*(-k,t)   \\
\end{pmatrix}
\end{gather}
is the $4\times4$ matrix in the Nambu$\times$Spin space with the Pauli matrices $\hat{\sigma}_{i}$ acting on the four-component wave function $\check{\psi}_{k}(t)$. 
The single particle matrix Hamiltonian in the spin space $\hat{H}(k,t)=\xi_k\hat{\sigma}_0-h(t){\bf z}_0\hat{{\bm \sigma}}+\alpha(\hat{{\bm \sigma}}\times  {\bf k}){\bf z}_0$ depends on the modulus $k=|{\bf k}|$ and the relative phase $\theta_k=\text{arg}(k_x+ik_y)$ of the momentum. Here $\xi_k= k^2/2m-E_F$ is a free particle spectrum measured from the Fermi level and $\alpha$ is a strength of RSOC. Hereafter we put $\hbar=1$. For simplicity we consider here the motion of QPs only in the $x-y$ plane neglecting their dispersion along the ${\bf z}_0$ axis, so that ${\bf k}=(k_x,k_y)$.
The pairing potential $\Delta(t)$ should satisfy the self-consistency equation, which at zero temperature $T=0$ can be written as follows
\begin{gather}\label{I_SCE}
\Delta(t)
=-\frac{\lambda}{2}\sum_\text{i.c.}\check{\psi}^{\dagger}_k(t) \check{\tau}_{\Delta}  \check{\psi}_k(t),
\end{gather}
where $\lambda$ is the pairing constant, $\check{\tau}_{\Delta}=(\hat{\tau}_x+i\hat{\tau}_y)\otimes i\hat{\sigma}_y/2$ and the independence of $\Delta$ on $\theta_k$ is taken into account. 
The summation here is performed over all solutions of  Eq. (\ref{Hamiltonian_BdG_SOI}) for different initial conditions (i.c.) at $t=0$. The  information about the dynamics as well as the distribution function of the QP excitations is contained in the functions $\check{\psi}_k(t)$, which self-consistently define the temporal evolution of the gap. 
In the homogeneous problem, the initial conditions are numbered by the momentum $k$, which, in the case of a spin-split superconductor, must be supplemented by the spin quantum number. All possible initial configurations of the QP states are defined by an equilibrium distribution function. The pairing potential $\Delta(t)$ can be chosen as a real function of time, and this choice will be justified below.

Generally speaking, the concept of an energy spectrum for a dynamical system is not clearly defined. However, in the case of adiabatic evolution one can introduce the eikonal approximation for the QP wavefunctions $\check{\psi}_{k}(t)\propto \check{\Psi}_{k}(t)e^{iS_{k}(t)}$, from which the adiabatic spectrum $E_{k}(t) = -\partial_t S_{k}$ can be extracted.
The functions $\check{\Psi}_{k}(t)$ are the instantaneous eigenstates of the Hamiltonian $\check{\mathcal{H}}(t)$ from Eq. (\ref{TDHam}).
The resulting spectrum is  
\begin{gather}\label{spectrum}
    E_{kn}(t) = \\ \notag
    \pm \sqrt{ E_0^2+\alpha^2k^2+h^2(t) \mp \text{sgn}(\sigma) 2\sqrt{\xi_k^2\alpha^2k^2+h^2(t)E_0^2} } 
\end{gather}
where $E_0=\sqrt{\xi_k^2+\Delta^2}$. We use the index $n \equiv \sigma \pm = \{\uparrow+, \downarrow+, \uparrow-, \downarrow-\}$ which refers to different spin subbands and positive/negative energy (these notations will be used in the text below).
There are four corresponding instantaneous eigenstates which can be written as
$\check{\Psi}_{k n}(t)=(u_{k \uparrow n}, u_{k \downarrow n}, v_{k \uparrow n}, v_{k \downarrow n})^T$. The detailed structure of the vectors is given in Appendix \ref{APP_A}.
The functions $\check{\Psi}_{kn}(t)$ form an orthonormal basis with he normalization condition $\check{\Psi}^{\dagger}_{kn}\check{\Psi}_{kn'}=\delta_{nn'}$ and the completness relation $\sum_{kn} \check{\Psi}_{kn}\check{\Psi}^{\dagger}_{kn} = \check{1}$.
Obviously, in the limit of the stationary Zeeman field,  $\check{\Psi}_{k n}$ becomes an exact solution of stationary problem (\ref{Hamiltonian_BdG_SOI}).  

It is important to keep in mind that in the presence of both RSOC and spin-splitting field the equilibrium gap value depends of the values of these fields $\Delta_\text{eq} = \Delta_\text{eq}(h, \alpha)$. In what follows, the RSOC strength $\alpha$ will be considered as a small parameter, and the dependence $\Delta(h, \alpha)$ will be neglected for simplicity.
Thus, the equilibrium gap value is defined as follows $$\Delta_\text{eq} = \Delta_0 = 2\hbar\omega_D e^{-\frac{1}{\lambda N(0) }},$$
where $\omega_D$ is Debye frequency and $N(0)$ is the density of states at Fermi energy.

\section{Linearized gap dynamics}\label{sec_3}

In this section we want to address the temporal evolution of a small fluctuation of the gap $\Delta_0+\delta\Delta(t)$ in the presence of the static spin-splitting field ${\bf h}=h_0{\bf z}_0$. The gap dynamics can be excited by some external pulse at $t=0$ or can be driven, for instance, by time-dependent spin-splitting field $\delta{\bf h}(t)=\delta h(t){\bf z}_0$.  In linear order in small perturbations $\delta\Delta(t),\delta h(t) \ll h_0 < \Delta_0$, the TDBdG equations for the QP wave functions read
\begin{gather} \label{III_tdproblem}
    i\frac{\partial}{\partial t}\check{\psi}_{k}(t) = \Big[ \check{\mathcal{H}_0 } + \check{ \mathcal{V}  }(t) \Big] \check{\psi}_{k}(t),
\end{gather}
where the operators in the Nambu$\times$Spin space are
\begin{gather}
\check{\mathcal{H}_0 }
=
\begin{pmatrix}
\hat{H}_0(k)   &    i\hat{\sigma}_y\Delta_0    \\
-i\hat{\sigma}_y\Delta_0  &     -\hat{H}^*_0(-k)   \\
\end{pmatrix}, 
\\ \notag
\check{ \mathcal{V} }(t)
=
\begin{pmatrix}
-\delta h(t)\hat{ \sigma}_z   &    i\hat{\sigma}_y \delta\Delta(t)    \\
-i\hat{\sigma}_y\delta\Delta(t)  & -\delta h(t)\hat{ \sigma}_z
\end{pmatrix},
\end{gather}
and single particle Hamiltonian is
$\hat{H}_0(k)=\xi_k\hat{\sigma}_0-h_0\hat{ \sigma}_z+\alpha(k_y\hat{\sigma}_x-k_x\hat{\sigma}_y)$.

Time-dependent equation (\ref{III_tdproblem}) can be written in the adiabatic basis using stationary eigenfunctions $\check{\Psi}_{kn}$ of the operator $\check{\mathcal{H}}_0$. Additionally, the RSOC energy $\alpha k\approx \alpha k_F$ is considered a perturbative parameter. By approximating the eigenvectors up to first order in $\alpha k_F/\Delta_0$ (see Appendix {\ref{APP_A}}), we can infer from equation (\ref{I_SCE}) that the fluctuation in the gap will have an order up to $\mathcal{O}(\alpha^2 k_F^2/\Delta_0^2)$. However, in the general case, the gap $\Delta$ should not be affected by the direction of the SOC. Therefore, the first-order change in the gap $\delta\Delta\propto \mathcal{O}(\alpha k_F/\Delta_0)$ must vanish.

Instead of the general eikonal theory, we use the perturbative approach with the ansatz written in terms of the dynamical phase
\begin{gather} \label{III_ansatz}
\check{\psi}_k(t)=\sum_n \check{\Psi}_{kn} C_{kn}(t)e^{-iE_{kn}t}.
\end{gather}
The index $n=\{\uparrow+, \downarrow+, \uparrow-, \downarrow-\}$  denotes the spectral branches and all negative/positive energy terms are involved into the dynamics of QPs. Substituting the function (\ref{III_ansatz}) into Eq. (\ref{III_tdproblem}) we obtain the equation for the dynamics of the coefficients
\begin{gather}\label{III_dCdt}
i\frac{\partial}{\partial t}C_{km}(t)=
\sum_n  \check{\Psi}_m^{\dagger}\check{\mathcal{V}}(t)\check{\Psi}_n e^{- i (E_n-E_m) t}C_{kn}(t),
\end{gather}
which completely determine the temporal evolution of gap $\Delta(t)$ through the self-consistency equation
\begin{gather}\label{III_gap_full}
\Delta_0 + \delta\Delta(t) = \\ \notag 
-\frac{\lambda}{2}\sum_\text{i.c.}\sum_{n,n'}C^*_{kn}(t)C_{kn'}(t)e^{-i(E_{n'}-E_n)t}\check{\Psi}_{kn}^{\dagger} \check{\tau}_{\Delta}  \check{\Psi}_{kn'}.
\end{gather}

In the case of zero temperature $T=0$ there are two possible initial configurations at $t=-\infty$: All QP states with energies below Fermi level in the first(second) spin subband with $\sigma=\uparrow(\downarrow)$ are fully occupied for all momenta with $\xi_k\in(-\omega_D, \omega_D)$. The corresponding initial conditions can be written as 
\begin{gather}\label{i_c_}
C_{kn}(t=-\infty) = \delta_{n,l} 
\end{gather}
with Kronecker delta $\delta_{n,n'}$ and indices $l=\{\uparrow-, \downarrow-\}$. Therefore, it is natural to linearize the equation (\ref{III_dCdt}) as follows
\begin{gather}\label{linearization}
    C_{kn}(t)=C_{kn}(-\infty)+\delta C_{kn}(t).
\end{gather}

Performing Laplace transform in the complex plane $s=i\omega+\zeta$ for the linearized equations (\ref{III_dCdt}, \ref{III_gap_full}, \ref{linearization}) (see Appendix \ref{APP_B}) we get the following expression for the gap perturbation
\begin{gather}\label{III_gap}
\delta\Delta(s) =\Big[ \mathcal{K}_0(s)+\mathcal{K}_+(s)+\mathcal{K}_-(s)  \Big] \delta\Delta(s) \\ \notag
+  \delta h(s) \Big[\mathcal{F}_+(s) - \mathcal{F}_-(s)\Big] + \mathcal{I}(s) .
\end{gather}
Here $\mathcal{K}_{0,\pm}(s)$ represents kernels of the self-consistency equation; $\mathcal{F}_\pm(s)$ defines the dynamical structure of the "force" term (in analogy with a mechanical oscillator) related with $\delta h(t)$; and $\mathcal{I}(s)$ includes all terms related to perturbations at the moment $t=0$. 
Due to the absence of particle-hole asymmetry, which couples the phase and amplitude fluctuations \cite{pekker_amplitudehiggs_2015}, the imaginary part of $\delta\Delta(s)$ naturally vanishes and we consider only amplitude (or Higgs) modes of the superconducting gap.
Knowing the function $\mathcal{K}(s)$ one can find eigenfrequencies and free dynamics of the system, while $\mathcal{F}_\pm(s)$ induces the driven dynamics. We will conduct a thorough examination of these terms below.

\subsection{Spin-split Higgs modes}

\begin{figure}[] 
\centering
\begin{minipage}[h]{0.95\linewidth}
\includegraphics[width=1.0\textwidth]{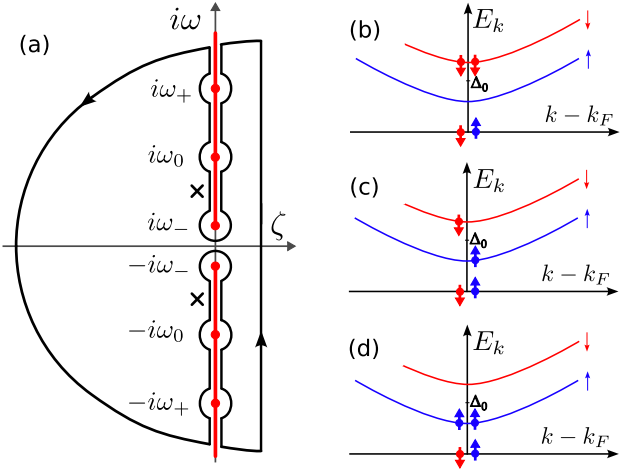} 
\end{minipage}
\caption{ {\small (a) Branch points $\omega_0=2\Delta_0, \omega_\pm = 2(\Delta_0\pm h_0)$ (red dots) corresponding to the kernels $\mathcal{K}_{0}(s)$ and $\mathcal{K}_\pm(s), \mathcal{F}_\pm(s)$ [Eq. (\ref{III_gap})] in the complex plane $s=\zeta+i\omega$. Red lines show the chosen branch cuts. Black crosses correspond to the poles of the external force $\delta h(s)$. (b-d) Illustration of physical mechanism behind the appearance of three eigenfrequencies $\omega_+$ (b), $\omega_0$ (c), $\omega_-$ (d).  }}
\label{fig1}
\end{figure}

It is known that in the absence of a spin-splitting field and RSOC the Higgs mode has a singular behavior in the vicinity of the eigenfrequency $\omega=2\Delta_0$, which defines the free evolution of the gap perturbation $\delta \Delta(t)\propto \cos(2\Delta_0t)/\sqrt{t}$ \cite{doi:10.1146/annurev-conmatphys-031119-050813}. 
Since the energy of the Higgs mode lies at the lower bound of the QP spectrum, the oscillatory behavior here can be represented as a coherent decay and formation of a Cooper pair into two QPs with opposite spins and energies $\Delta_0$ at $k\approx k_F$. The contribution from the pairs of QPs with other momenta leads to the inhomogeneous broadening of the mode with the corresponding damping law.
The presence of Zeeman field and RSOC makes the dynamics more complicated. 
To analyze the eigenmodes of the superconductor one can set $\delta h(t)=0$ and write the self-consistency equation as follows 
$$ \chi_{\Delta \Delta}^{-1}(s)\delta\Delta(s) = 0, $$
where we define the bare pair susceptibility 
\begin{gather}\label{chi_D}
\chi_{\Delta \Delta}(s)=\frac{1}{1-\mathcal{K}_0(s)-\mathcal{K}_+(s)-\mathcal{K}_-(s)}.
\end{gather}
The corresponding kernels read (see Appendix \ref{APP_B})
\begin{gather}\label{kern_nat}
\mathcal{K}_0(s) = \Big\langle \frac{2\xi^2}{E_0}\frac{1}{s^2+4E^2_0} \Big\rangle \propto \mathcal{O}\Big(\frac{\alpha^0 k^0_F}{\Delta_0^0}\Big), \\ \notag
\mathcal{K}_{\pm}(s) = \Big\langle \mathcal{A}^2(\xi)\frac{E_0\pm h_0}{s^2+4(E_0\pm h_0)^2} \Big\rangle \propto \mathcal{O}\Big(\frac{\alpha^2 k^2_F}{\Delta_0^2}\Big),
\end{gather}
where the notation $\big\langle \dots \big\rangle = \lambda N(0) \int_{-\omega_D}^{\omega_D}d\xi$ is used. 
The function $\mathcal{A}(\xi)\propto \check{\Psi}^{0\dagger}_{kn}\check{\tau}_{\Delta}\check{\Psi}^{0}_{kn}\propto\alpha k_F/\Delta_0$ is proportional to nonzero triplet component of the wave function, therefore the kernels $\mathcal{K}_\pm$ are of the second order in the RSOC parameter. 

The frequencies of the eigenmodes of the superconducting condensate can be traced out from the condition $|\chi_{\Delta \Delta}^{-1}(\omega)|=0$, which reflects the singular points of the kernels (\ref{kern_nat}). Consider these points in more detail. Instead of straightforward integrating, we are going to implement the analysis in the spirit of the work \cite{1974JETP38.1018V} and analytically obtain the limit ${\zeta \rightarrow 0}$. 
The functions ${\mathcal{K}_{0,\pm}(s\rightarrow \omega)}$ can be represented as ${\mathcal{K}(s)=\mathcal{K}'(\omega)+i\text{sgn}(\omega\zeta) \mathcal{K}''(\omega)}$.
The real parts of the kernels 
\begin{align}
\frac{\mathcal{K}_0'(\omega)}{\lambda  N(0)} &=  \fint_{-\omega_D}^{\omega_D}  \frac{2\xi^2}{\sqrt{\xi^2+\Delta_0^2}(4\xi^2+4\Delta_0^2-\omega^2)} d\xi, \\ \label{Kern_1}
\frac{\mathcal{K}_\pm'(\omega)}{\lambda  N(0) } &= \fint_{-\omega_D}^{\omega_D}  \frac{\mathcal{A}^2(\xi) (E_0\pm h_0)}{4(E_0-h_0)^2\pm |\omega|^2} d\xi,
\end{align}
are regular on the imaginary axis $s=i\omega$.
The imaginary parts are
\begin{align}
    \frac{\mathcal{K}_0''(\omega)}{\lambda N(0)} &= -\frac{\pi}{2}\frac{\sqrt{\omega^2-\omega_0^2}}{|\omega|}\Theta[\omega^2-\omega_0^2], \\ \label{Kern_2}
    \frac{\mathcal{K}_{\pm}''(\omega)}{\lambda N(0)} &= 
    -\frac{\pi}{8} \frac{|\omega| \mp 2h_0 }{\xi_\pm}   \mathcal{A}^2(\xi_\pm) \Theta[\omega^2-\omega_\pm^2],  
\end{align}
where $\xi_\pm=\frac{1}{2}\sqrt{\big( |\omega| - \omega_\pm \big)^2+4\Delta_0(|\omega|-\omega_\pm)}$. 
The discontinuities at the real axis $\zeta$ mean the existence of the branch points 
\begin{gather}\label{III_freq} 
    \omega_0 = 2\Delta_0, \\ \notag
    \omega_+ = 2(\Delta_0+ h_0), \\ \notag
    \omega_- = 2(\Delta_0- h_0),
\end{gather}
and corresponding cuts in the complex plane [Fig. \ref{fig1}(a)].  

The analysis of the  general linear response of the order parameter can be significantly simplified by expanding the susceptibility $|\chi_{\Delta\Delta}(\omega)|$ in the powers of the small parameter $\alpha k_F/\Delta_0$, since the kernels $\mathcal{K}_{\pm}\propto \mathcal{O}(\alpha^2 k^2_F/\Delta_0^2)$.
As mentioned before, the maximum order we can take into account is $|\chi_{\Delta\Delta}|\propto\mathcal{O}(\alpha^2 k_F^2/\Delta_0^2)$. 
The resonance condition $|\chi_{\Delta \Delta}^{-1}(\omega)|=0$ is satisfied at $\omega=\omega_\pm$ where the kernels $\mathcal{K}_\pm''(s)$ have a singularity (note that $\mathcal{A}$ is regular at $\xi=\xi_\pm$), and at $\omega=\omega_0$, where the function $\mathcal{K}_0''(s)$ goes to zero.
Thus, the branch points (\ref{III_freq}) define new eigenmodes of the superconductor in the presence of spin-splitting field and weak RSOC. 
By taking some constant initial condition $\mathcal{I}(s)=const$ in the RHS of Eq. (\ref{III_gap}) and using inverse Laplace transform one can consider an impulse response of the superconductor. It can be shown that the peculiarities in the vicinities of the eigenfrequencies lead to three partial contribution to the long-time gap dynamics 
\begin{gather}\label{free_higgs}
\delta\Delta(t)\propto \frac{\cos(\omega_0t+\pi/4)}{\sqrt{\Delta_0t}} \\ \notag
+\Big(\frac{\alpha k_F}{\Delta_0}\Big)^2 A_+(h_0) \frac{\cos(\omega_+t+\pi/4)}{\sqrt{\Delta_0t}}  \\ \notag 
+\Big(\frac{\alpha k_F}{\Delta_0}\Big)^2 A_-(h_0) \frac{\cos(\omega_-t+\pi/4)}{\sqrt{\Delta_0t}}
\end{gather}
with some amplitudes $A_\pm(h_0)$, which can be identified as spin-split Higgs modes.
Details of the derivation of $\delta\Delta(t)$ from Eq. (\ref{III_gap}) will be provided in the next subsection together with Appendix \ref{APP_C}.

Appearance of the frequencies (\ref{III_freq}) and corresponding oscillations (\ref{free_higgs}) in the spin-split superconductor can be explained qualitatively. Coherent decay of the Cooper pairs from the Fermi level can occur into two different spin subbands of the QP spectrum. When two electrons with opposite spins from a pair dissociate into two QP at $k\approx k_F$ with the energies $\Delta_0\pm h_0$ without spin-flipping, then the total decay energy is equal to QP threshold $\approx 2\Delta_0$. This process corresponds to the mode $2\Delta_0$ and shown in Fig. \ref{fig1}(c). A decay into two QPs with the same spins is possible in the presence of RSOC due to the effective spin-flip scattering. The energies of such two QPs are either $\Delta_0+ h_0$ or $\Delta_0 - h_0$. This process leads to the modes $2(\Delta_0\pm h_0)$ correspondingly [Fig. \ref{fig1}(b,d)]. 
Note that this naive interpretation of the complicated QP dynamics is valid for the sufficiently small RSOC $\alpha k_F\ll \Delta_0$. 

\begin{figure}[] 
\begin{minipage}[h]{0.9\linewidth}
\includegraphics[width=1.0\textwidth]{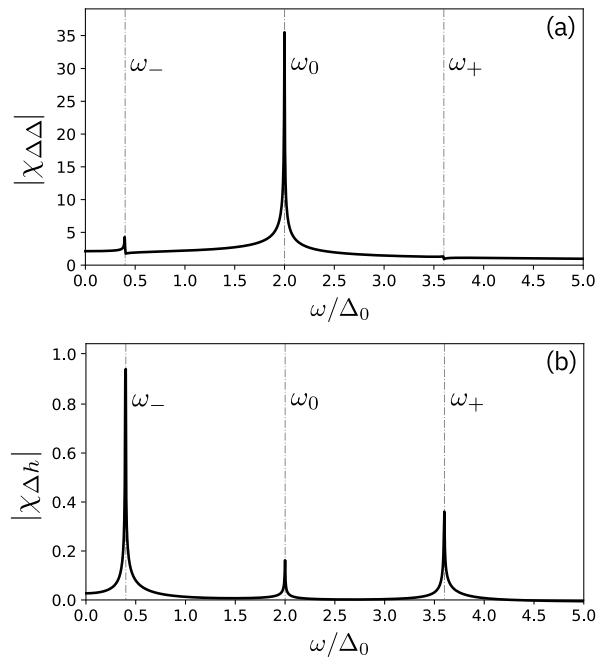} 
\end{minipage}
\caption{ {\small 
(a) The susceptibility $|\chi_{\Delta\Delta}(\omega)|$ of free dynamics of the gap perturbation $\delta\Delta(\omega)$. Features at the frequencies $\omega_{0}=2\Delta_0$ and $\omega_\pm=2(\Delta_0\pm h_0)$ correspond to Higgs modes resonances. The finite width of the resonances is due to the small imaginary part of the frequency $\omega+i\beta$.  
(b) Frequency dependence of the response functions $|\chi_{\Delta h}(\omega)|$ of the driven gap oscillations $\delta \Delta(t)$ excited by the Zeeman field $\delta h(t)=\delta h_0 e^{-\beta t} \cos(\omega t)$  (see Eq. \ref{dD}). Both plots are symmetrical with respect to $\omega \rightarrow -\omega$ and have the parameters $h_0=0.8\Delta_0$, $\alpha k_F=0.05\Delta_0$ and $\beta=0.002\Delta_0$.}}
\label{fig2}
\end{figure}

Numerically calculated susceptibility $|\chi_{\Delta\Delta}(\omega)|$ from Eqs. (\ref{chi_D}-\ref{kern_nat}) is shown in Fig. (\ref{fig2})(a). The observed resonances have a different parametric order of smallness. The Higgs mode with the frequency $\omega_0$ which exists in the absent the RCOS becomes dominating with more pronounced peak $|\chi_{\Delta\Delta}(\omega\approx \omega_0)|\propto \alpha^0 k_F^0$, whereas two other modes at shifted frequencies $\omega_\pm$ are of the order of $|\chi_{\Delta\Delta}(\omega\approx \omega_\pm)|\propto \alpha^2 k^2_F/\Delta_0^2$. These modes merge with $\omega_0$ at $h_0\rightarrow 0$ and disappear for $\alpha \rightarrow 0$.
It is expected that the excitation of the bare response of the superconductor can be implemented with the standard THz laser pump-probe techniques. The electric field of the pump pulse produces a quench of the spin-split superconductor and subsequent probe pulse detects the multifrequency Higgs oscillations.

Note, that a similar dynamics of the order parameter has been studied in the spin-orbit coupled Fermi gases \cite{behrle2018higgs,Wang_2015,dong2015dynamical}. In particular, the existence of the Higgs modes modified by the Zeeman field in the presence of strong SOC with $\alpha k_F \sim h(t) \sim E_F$ was discussed in Ref. \cite{Fan:52502}. The authors performed a numerical simulation of the one-dimensional Fermi superfluid and examined the excitation of the gap oscillations with few frequencies by abrupt change of the Zeeman field. Despite the significant differences between the models, there is a general tendency for the influence of the shift of spectral QP branches on the behavior of the order parameter modes.

\subsection{Coupling of Higgs modes and Zeeman field }

We found that, in addition to an electromagnetic field, the gap dynamics in a spin-split superconductor can be excited by a nonstationary component of Zeeman field ${\bf h}(t)=( h_0+\delta h(t)){\bf z}_0$. In this particular configuration the perturbation of the spin-splitting field $\delta h(t)$ appears in the self-consistency equation (\ref{III_gap}) in the first order, which is the trace of a dot product $({\bf h}_0 \cdot \delta{\bf h})$.
Note that the field $\delta h(s)$ is weighted by the functions 
\begin{gather}\label{F_force}
\mathcal{F}_\pm(s) = \Big\langle \mathcal{A}(\xi)\mathcal{B}(\xi)\frac{(E_0\pm h_0)}{s^2+4(E_0\pm h_0)^2} \Big\rangle \propto \mathcal{O}\Big(\frac{\alpha^2 k^2_F}{\Delta_0^2}\Big), 
\end{gather}
which can be written as $\mathcal{F}(s)=\mathcal{F}'(\omega)+i\text{sgn}(\omega\zeta) \mathcal{F}''(\omega)$ and have the same order in $\alpha k_F$ and the same analytical properties as the kernels $\mathcal{K}_\pm(s)$ in (\ref{Kern_1},\ref{Kern_2}), because both functions $\mathcal{A}^2(\xi)$ and $\mathcal{A}(\xi)\mathcal{B}(\xi)$ are regular for $\xi \in (-\omega_D, \omega_D)$.
The presence of the singular points in the force term makes the analysis of Eq. (\ref{III_gap}) more sophisticated, despite the fact that these points are shared with other kernels.

Consider the general case of forced oscillations of the order parameter driven by some field $\delta h(t)$ which is abruptly turned on at $t=0$.
Since we want to consider dynamical effects related only to the external force, we neglect the initial conditions $\mathcal{I}(s)$ in Eq. (\ref{III_gap}), e.g. assume the equilibrium system with $\delta\Delta(t)=0$. 
It is convenient to introduce the linear response function
\begin{gather}\label{chi} 
\chi_{\Delta h}(s) = \frac{ \mathcal{F}_+(s)-\mathcal{F}_-(s)}{ 1- \mathcal{K}_0(s)-\mathcal{K}_+(s)-\mathcal{K}_-(s) }
\end{gather}
and write the self-consistency equation as
${\delta\Delta(s) =\chi_{\Delta h}(s)\delta h(s)}$.
One can obtain the expression for $\delta \Delta(t)$ in the interval $t\in [0, \infty)$ using inverse Laplace method:
\begin{gather}
\delta\Delta(t) = \frac{1}{2\pi i} \int_{-i\infty+\epsilon}^{i\infty+\epsilon} 
\chi_{\Delta h}(s) \delta h(s) e^{st}ds,  
\end{gather}
where $\epsilon $ should be larger then the real part of the poles of $\delta h(s)$.
The integral can be evaluated using closed contour shown in Fig. \ref{fig1}(a). Making sure that all integrals on infinitely large and small arcs vanish and applying residue theorem we get
\begin{gather}\label{III_integral}
\delta\Delta(t) = \sum_p  \chi_{\Delta h}(s_p) e^{s_pt} \underset{s=s_p}{\text{Res}} \Big[\delta h(s) \Big]  \\ \notag
+\frac{2}{\pi}\int_{\omega_-}^{\infty} 
\text{Im}\chi_{\Delta h}(s)\big|_{\zeta\rightarrow +0}
\text{Im}\Big[ e^{i\omega t}  \delta h(i\omega)  \Big]
d\omega. 
\end{gather}
The first term represents the contribution from the poles of the external field $\delta h(t)$, while the second term is the contribution from the integrals along the branch cuts.
%

The imaginary part of the susceptibility ($\ref{chi}$) can be expanded up to the second order in $\alpha k_F/\Delta_0$, since all the kernels $\mathcal{F}_{\pm}, \mathcal{K}_\pm \propto\mathcal{O}(\alpha^2 k^2_F/\Delta_0^2)$. That allows one to distinguish different strongly dominant terms of the function $\text{Im}\chi_{\Delta h}(\omega)$ in (\ref{III_integral}) in the vicinity of different branch points (\ref{III_freq}) and estimate their contribution to an asymptotic expression for $\delta \Delta(t)$.
The detailed calculations are provided in Appendix \ref{APP_C}. Here we write the result for the superconducting gap oscillations, which at large times $h_0^{-1} \ll t$ reads
\begin{widetext}
\begin{gather}\label{dD}
\delta\Delta(t) \approx
 \sum_p  \chi_{\Delta h}(s_p) e^{s_p t} \underset{s=s_p}{\text{Res}} \big[ \delta h(s)\big]
+  \frac{4 \Delta_0 }{ \pi\sqrt{\pi}} 
\frac{ \big[\mathcal{F}_+'(\omega_0)-\mathcal{F}_-'(\omega_0)\big]}{\lambda N(0)}
\frac{ \text{Im} \Big[  \delta h(i\omega_0) e^{i(\omega_0 t + \pi/4)}   \Big] }{ \sqrt{ \Delta_0 t}} 
\\ \notag
+
\frac{\sqrt{\pi} }{8\sqrt2}
\frac{(\alpha k_F)^2\Delta_0}{(\Delta_0-h_0)^2}
\sum_{j=\pm}
\frac{\lambda N(0) \big[1- \mathcal{K}_0'(\omega_j) \big]}{ \big[1- \mathcal{K}_0'(\omega_j) \big]^2 + \big[ \mathcal{K}_0''(\omega_j) \big]^2 }
\frac{\text{Im} \Big[  \delta h (i\omega_j)  e^{i(\omega_j t + \pi/4)}   \Big] }{ \sqrt{\Delta_0 t}}.
\end{gather}
\end{widetext}
The first term here is related to the forced oscillations of the gap, caused by the Zeeman field $\delta h(t)$. The last three terms correspond to the free oscillations triggered by $\delta  h(t)$ at $t=0$ in the long time asymptote, with three characteristic frequencies (\ref{III_freq}) and square-root damping law. The latter can be interpreted as partial contribution from the  Higgs modes in the spin-splitting field $h_0$.

The eigenmodes decay at $t \rightarrow \infty$ and in the long-time asymptote the forced oscillations prevail.
Consider the steady-state behavior of $\delta\Delta(t)$ (the first term in Eq. (\ref{dD})) in the time interval restricted by the inelastic relaxation processes where the presented description of the coherent gap dynamics is valid. Assume the general harmonic perturbation $\delta h(t)=\delta h_0 e^{-\beta t} \cos(\omega t)$ at $t\geq 0$  with small finite damping factor $\beta\rightarrow 0$.The amplitude of the driven gap perturbation $\delta\Delta(t)$ is defined by the Zeeman field $h_0$ and susceptibility $\chi_{\Delta h}(s_0)$ taken at the pole of the force $s_0=-\beta+i\omega$. The numerically integrated shape of $|\chi_{\Delta h}(\omega)|$ is shown in Fig. (\ref{fig2})(a). The response of the superconductor, as expected, has three resonance peaks at the frequencies $\omega_{0,\pm}$. However, since the the external field $\delta{\bf h}(t)$ couples to the gap through the RSOC, the amplitude of the susceptibility in the vicinity of the resonances has the same order of smallness $|\chi_{\Delta h}(\omega_{0,\pm})|\propto \mathcal{O}(\alpha^2 k_F^2/\Delta_0^2)$, which differs from the bare response (\ref{chi_D}).

In this section, we have solely focused on the longitudinal component of the field perturbation $\delta h(t) {\bf z}_0$ with respect to the stationary field $h_0{\bf z}_0$. However, it is also possible to introduce the time-dependent transversal component $\delta {\bf h}\bot(t)$ and examine its dynamic interaction with the superconducting system in Eq. (\ref{III_tdproblem}). This component generates triplet correlations, but these do not contribute to the order parameter since only singlet pairing in (\ref{I_SCE}) is considered. Consequently, in the second-order perturbation theory with respect to $\alpha k_F/\Delta_0$, there is no linear coupling between the field $\delta {\bf h}\bot(t)$ and the gap $\delta\Delta(t).$ This outcome is unsurprising since the only true scalar in this regime $(\delta {\bf h}_\bot \cdot {\bf h}_0)$ is zero.

\section{Evolution of QP states in strong Zeeman field}\label{sec_4}

\begin{figure*}[] 
\centering
\begin{minipage}[h]{0.9\linewidth}
\includegraphics[width=1.0\textwidth]{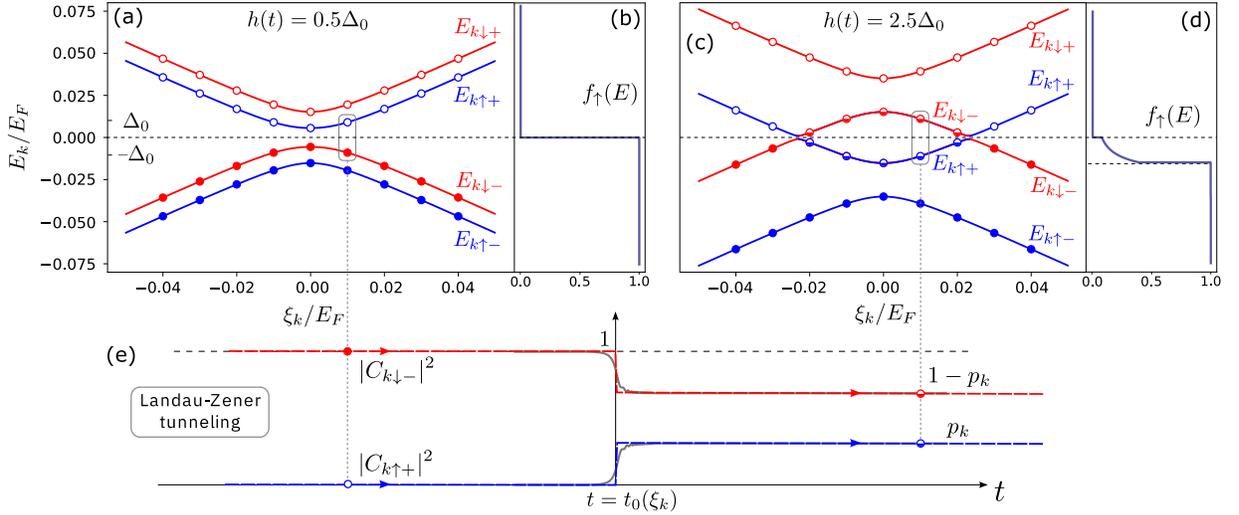} 
\end{minipage}
\caption{ {\small 
(a,b) QP spectrum $E_k$ from (\ref{spectrum}) for two values of Zeeman field $h(t)$ before (a) and after (b) avoided crossing. Colored/empty circles correspond to filled/empty states. The parameters are chosen as follows: $\Delta_0/E_F=0.01$, $\alpha/E_F=0.0025$. In (e) the schematic temporal evolution of the filling probabilities $|C_k|^2(t)$ for two states at fixed $\xi_k$ is shown. The gray lines show a tunneling process similar to the real one in the vicinity of the avoided transition point
$t_0(\xi_k)=\sqrt{\xi_k^2+\Delta_0^2}/\gamma$, while red and blue lines refer to transition matrix approximation of the LZSM tunneling with the probability $p_k$.
(b, d) The QP distribution function for one spin projection $f_\uparrow(E,t)$ from Eq. (\ref{distrib}) before and after crossing of spectral branches at $\delta_\text{LZ}=0.5$.
 }}
\label{fig3}
\end{figure*}

In this section we address the case of a linearly growing spin-splitting field $h(t)=\gamma t$, which can exceed the equilibrium value of the superconducting gap $\Delta_\text{eq}\equiv \Delta_0$ and thus provide the crossing of the two QP spectral branches $E_{\uparrow+}(\xi_k)$ and $E_{\downarrow-}(\xi_k)$ from different spin subbands [Fig. \ref{fig3}(a, c)]. In the collisionless regime and in the absence of RSOC the intersecting spectral branches do not interact, so that the occupation of the quasiparticle states defined at $t=-\infty$ does not change in time.
This means that the self-consistent gap function will not change even above the paramagnetic limit $h(t)>\Delta_0$ and will be defined by the initial condition $\Delta(t)=\Delta_0$. It is clear from the general assumptions that the spin-orbit coupling is capable of provoking the interplay between QP states with different spins, and we investigate the mechanism of such an interaction and the effect on the superconducting order parameter $\Delta(t)$. As mentioned in the Section \ref{sec_2}, we will treat the RSOC energy  as a small parameter $\alpha k_F/\Delta\ll 1$. Therefore, we neglect the dependence of the equilibrium gap $\Delta_\text{eq}$ on $\alpha$ and assume $\Delta_\text{eq} \equiv \Delta_0$.

\subsection{Adiabatic evolution of QP states}\label{sec_4_A}

The evolution of QP wavefunction of the TDBdG equations (\ref{Hamiltonian_BdG_SOI}) can be regarded with the help of general adiabatic ansatz
\begin{gather}\label{wf}
\check{\psi}_k(t)=\sum_nC_{kn}(t)\check{\Psi}_{kn}(t),
\end{gather}
where $\check{\Psi}_{kn}(t)$ are the instantaneous eigenstates of the Hamiltonian (\ref{TDHam}).
Here all negative/positive energy terms with the indices $n=\{\uparrow +, \downarrow+, \uparrow -, \downarrow-\}$ are taken into account. The coefficients $C_{kn}(t)$ define the occupation of QP states and its temporal evolution. The initial conditions for $C(t)$ should be fixed by the equilibrium distribution at $t=0$. In the case of spin-split superconductor at zero temperature $T=0$ there are two possible initial configurations: All QP states with energies below Fermi level in the first(second) spin subband with $\sigma=\uparrow(\downarrow)$ are fully occupied for all momenta with $\xi_k\in(-\omega_D, \omega_D)$.
With short notations one can write this as $C_{kn}(t=0) = \delta_{n,l}$, where $\delta_{n,n'}$ is Kronecker delta and $l=\{\uparrow-, \downarrow-\}$. This means that for the given $l$ we have $C_{kl}(t=0)=1$, and all other $C_{k(n\neq l)}(t=0)=0$. 

The vector 
\begin{gather}\label{C_vec}
\hat{C}_{k}(t)=( C_{k\uparrow+}, C_{k\downarrow+}, C_{k\uparrow-}, C_{k\downarrow-} )^T
\end{gather}
contains all the information about the dynamics of the QP states. Corresponding adiabatic temporal evolution can be described with the help of the unitary operator $\hat{C}_{k}(t_2)=\hat{U}_k(t_2,t_1)\hat{C}_{k}(t_1)$ where $\hat{U}_{k}=\text{diag}(U_{k\uparrow+}, U_{k\downarrow+}, U_{k\uparrow-}, U_{k\downarrow-})$ and
\begin{gather} \label{U_t}
U_{kn}(t_2, t_1)=\exp\Big(-i\int_{t_1}^{t_2}E_{kn}(t)dt\Big).
\end{gather}
The interaction of the branches $E_{k\uparrow +}(t)$ and $E_{k\downarrow -}(t)$ leads to \textit{avoided crossing} of the QP levels at fixed energy $\xi_k$ with the splitting proportional to $\alpha k_F$. Thus the adiabatic approximation is justified only for the levels with $E_k(t) \gg \alpha k_F$, e.g. far enough from the crossing points. 
Therefore, for the Zeeman field $h(t)\lesssim \Delta_0$ all nonadiabatic transitions are suppressed and the gap function defined by the self-consistency equation (\ref{I_SCE}) is equal to the equilibrium value $\Delta(t)=\Delta_0$.

\subsection{Transition evolution matrix}\label{sec_4_B}

The avoided crossing between the spectral terms at $h(t)\gtrsim\Delta_0$ should be described in terms of nonadiabatic dynamics. 
For this we consider the branch intersection as consecutive avoided crossing of pairs of the QP states with fixed energy $\xi_k$ at the time instant $t_0(\xi_k)=\sqrt{\xi_k^2+\Delta^2}/\gamma$ [Fig. \ref{fig3}]. 
For each crossing at $\xi_k\in (-\omega_D, \omega_D)$ it is possible to formulate the time-dependent Landau-Zener-St\"{u}ckelberg-Majorana (LZSM) problem \cite{ivakhnenko_nonadiabatic_2023}, which describes the transitions between two QP states with different spins during their temporal evolution. Note that resulting nonadiabatic tunneling is equivalent to dynamical spin-flip process. 

In general, the description of such a tunneling (or LZSM problem) requires joint solution of TDBdG equation (\ref{Hamiltonian_BdG_SOI}) and self-consistency equation (\ref{I_SCE}). 
However, some important results can be obtained analytically using certain approximations:

(i) If the time variation of the gap function $\Delta(t)$ is small on the typical tunneling time scale $\tau_\text{LZ}$ (see Appendix \ref{APP_D}), then the tunneling of QP states is not affected by the dynamics of the order parameter. 

(ii) On the other hand, the gap $\Delta(t)$ is defined by all states in range $\xi_k\in(-\omega_D, \omega_D)$, and a time-dependent perturbation of the states caused by the dynamical LZSM transition makes a small contribution to the sum over all $\xi_k$. Thus, one can neglect the transient dynamics of the coefficients $\hat{C}_k(t)$ in the vicinity of a transition point for each $\xi_k-$th mode. This also means that one can investigate the tunneling problem with the help of so-called transition evolution matrix \cite{ivakhnenko_nonadiabatic_2023} connecting two adiabatic regimes before ($t<t_0-$) and after ($t>t_0+$) avoided crossing [Fig. \ref{fig3}(e)]. These conditions allow one to effectively decouple the LZSM problem from the self-consistency equation and solve them without self-consistency.

Taking into account all these assumptions, the time evolution of the vector $\hat{C}_{k}(t)$ from the adiabatic ansatz (\ref{wf}) is described as follows
\begin{equation}\label{qp_evol}
    \hat{C}_{k}(t)=
\left\{ \begin{aligned} 
    \hat{U}_k(t,t_0+)\hat{S}_{\text{LZ}}\hat{U}_k(t_0-,0)\hat{C}_{k}(0), & \quad t>t_0(\xi_k) \\
    \hat{U}_k(t,0)\hat{C}_{k}(0), & \quad t<t_0(\xi_k) 
\end{aligned} \right. .
\end{equation}
Here the nonadiabatic transitions between QP states are included into transition matrix $\hat{S}_{\text{LZ}}$, which acts on the state vector $\hat{C}_{k}(t)$ at the time instant $t=t_0(\xi_k)$. 
The matrix $\hat{S}_{\text{LZ}}$ can be obtained by considering the interaction of two intersecting energy branches $E_{\uparrow+}$ and $E_{\downarrow-}$ in the TDBdG equation (\ref{Hamiltonian_BdG_SOI}). Using so-called diabatic basis (basis of Hamiltonian (\ref{TDHam}) in the absence of RSOC) one gets a system of dynamical equations, the asymptotic solution of which forms a transition matrix describing the passage through the avoided intersection point. Then we go to the original adiabatic basis (\ref{wf}) and get the matrix $\hat{S}_\text{LZ}$.
The complete derivation of $\hat{S}_{\text{LZ}}$ is presented in Appendix \ref{APP_D} and it reads
\begin{gather} \label{S_lz}
\hat{S}_{\text{LZ}}=
\begin{pmatrix}
\sqrt{p_k} & 0 & 0 &  \sqrt{1-p_k}e^{i(\dots)} \\
0 & 1 & 0 & 0 \\
0 & 0 & 1 & 0 \\
-\sqrt{1-p_k}e^{-i(\dots)} & 0 & 0 & \sqrt{p_k}     
\end{pmatrix}, 
\end{gather}
where
$(\dots)=\chi_k-\theta_k-\frac{\pi}{2}\text{sgn}(\alpha)$.
The coefficient $$p_k=\exp\Big[-\delta_{\text{LZ}}\frac{\Delta^2}{\xi_k^2+\Delta^2}\Big]$$ is expressed through the dimensionless LZSM parameter $\delta_\text{LZ}=\pi\alpha^2k_F^2/\gamma$ and determines the probability of tunneling between QP states with different spins. The transition is accompanied by the appearance of the Stokes phase $\chi_k$ (see Appendix \ref{APP_D}) and the phase $\theta_k=\arg\big(k_x+ik_y\big)$. 

To avoid confusion, we use the same notations for the spectral branches (\ref{spectrum}) before ($t<t_0-$) and after ($t>t_0+$) QP transitions, as shown in Fig. \ref{fig3}. Thereby, we do not need to keep track of the indices of the eigenvectors $\check{\Psi}_{kn}(t)$ and the evolution operators $U_{kn}(t)$ from (\ref{U_t}). It is sufficient that these functions take into account the permutation of the branches of the spectrum (\ref{spectrum}), so that all QP levels change their indices after the transition in accordance with the chosen notation.
%
%

\subsection{Time dependence of superconducting gap}\label{sec_4_C}

The time-dependent order parameter subjected to the field $h(t)\gtrsim\Delta_0$ depends on both the adiabatic wave function (\ref{wf}) and nonadiabatic LZSM tunneling (\ref{qp_evol}). The calculation of $\Delta(t)$ can be accomplished using the self-consistent equation (\ref{I_SCE}), which gets the following form
\begin{gather}\label{g_1}
\Delta(t)=-\frac{\lambda }{2}\sum_{l}\sum_k\sum_{n,n'}C_{kn}^*(t)C_{kn'}(t)\check{\Psi}_{kn}^{\dagger}\check{\tau}_{\Delta}\check{\Psi}_{kn'},
\end{gather}
where index $l$ means different initial configurations of the occupation of the QP spectrum at $t=0$ (see Section \ref{sec_4_A}). 
The first configuration with $C_{kn}(t=0)=\delta_{n,\uparrow-}$ corresponds to occupation of all QP states belonging to the spectral branch $E_{k,\uparrow-}$ for all momenta with $\xi_k\in(-\omega_D, \omega_D)$. 
The evolution of the coefficients $C_{kn}(t)$ is determined by the Eq. (\ref{qp_evol}) together with Eq. (\ref{C_vec}-\ref{U_t}).
Since the branch $E_{k,\uparrow-}$ does not cross with other branches, the coefficients $C_{kn}(t)$ have a trivial adiabatic dynamics, which can be written as follows
\begin{gather}
\hat{C}_{k}(t)
=
\begin{pmatrix}\label{C_0}
0       \\
0       \\
U_{\uparrow-}\big(t,0\big)   \\
0   
\end{pmatrix}.
\end{gather}
The second initial configuration with $C_{kn}(t=0)=\delta_{n,\downarrow-}$ leads to the intersection of the filled branch $E_{k,\downarrow-}$ and empty branch $E_{k,\uparrow+}$. Using equation (\ref{qp_evol}) we obtain a nontrivial dynamics of the states with LZSM tunneling, which reads
\begin{gather}\label{C_2}
\hat{C}_{k}(t)   = \\ \notag
\begin{pmatrix}
\sqrt{1-p_k}e^{i(\dots)} U_{\uparrow+}\big(t,t_0+\big) U_{\downarrow-}\big(t_0-,0\big) \Theta\big[t-t_0\big] \\
0       \\
0      \\
\Big(\sqrt{p_k}\Theta\big[t-t_0\big]+\Theta\big[t_0-t\big]\Big) U_{\downarrow-}\big(t,0\big)    
\end{pmatrix}.
\end{gather}
Here $(\dots) =\chi_k-\theta_k-\frac{\pi}{2}\text{sgn}(\alpha)$ and $\Theta(t)$ is the Heaviside function.

Substituting coefficients (\ref{C_0}) and (\ref{C_2}) obtained from different initial conditions together with the QP wavefunctions $\hat{\Psi}_{kn}$ from (\ref{APP_A_eigenf}) into the self-consistency equation (\ref{g_1}) we get
\begin{gather}\label{D_h}
\Delta(t)=\lambda  \sum_{ |\xi_k|>\sqrt{h^2-\Delta^2} }  u_0v_0
\\ \notag
+\lambda \sum_{ |\xi_k|<\sqrt{h^2-\Delta^2}} 
\Bigg[    
 \Big(\overset{=1}{|C_{k\uparrow-}|^2} 
    +\overset{\propto p_k}{|C_{k\downarrow-}|^2}
    -\overset{\propto 1-p_k}{|C_{k\uparrow+}|^2}
      \Big) \frac{u_0v_0}{2} \\ \notag
     + 
    u_0u_1 ie^{-i\theta_k} C_{k\uparrow+}^*C_{k\downarrow-}
    +v_0v_1 (-i)e^{i\theta_k} C_{k\downarrow-}^*C_{k\uparrow+}\Bigg].
\end{gather}
The last two terms are of the order of $\mathcal{O}(\alpha k_F/\Delta)$, so it is convenient to write the gap function as follows
\begin{gather}\label{d_h+dd}
\Delta(t)=\Delta_h[h(t)]+  \delta\Delta(t).
\end{gather} 
We have identified two contributions that have significantly different origins: $\Delta_h$ is defined by the amplitude of the LZSM tunneling and depends on time only through the Zeeman field $h(t)$; 
$\delta\Delta(t)\propto \mathcal{O}(\alpha k_F/\Delta)$ is defined by cross-terms and reflects interference effects between QP wavefunctions caused by LZSM transitions and depends on time explicitly. 

\begin{figure}[] 
\centering
\begin{minipage}[h]{1.0\linewidth}
\includegraphics[width=1.0\textwidth]{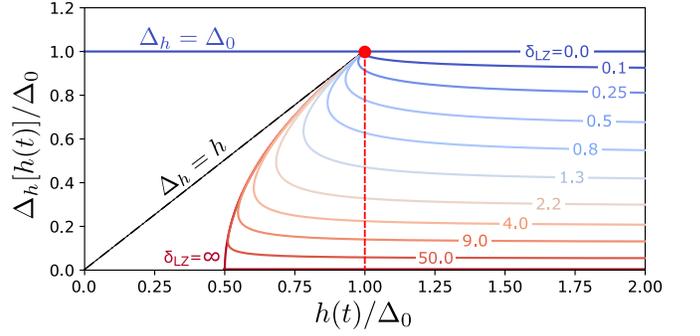} 
\end{minipage}
\caption{ {\small Functional dependence of the superconducting gap $\Delta_h$ on the spin-splitting field from expression (\ref{D_h(t)}) for different values of $\delta_\text{LZ}$. The dashed-dotted line separates two regions where $\Delta_h\lessgtr h$. 
The dashed line shows the critical value of the field $h(t)=\Delta_0$
and the red circle marks the point of change in the behavior of the gap $\Delta_h$ in the region $\Delta_h>h$. After this point the equilibrium solution $\Delta_h=\Delta_0$ should jump to one of the solutions fixed by the parameter $\delta_\text{LZ}$.  }}
\label{fig4}
\end{figure}

If one neglects the small perturbation $\delta\Delta(t)$ in (\ref{d_h+dd}) then it becomes possible to get a simplified self-consistency equation for $\Delta_h[h(t)]$ from Eq. (\ref{D_h}). In general form it reads
\begin{gather}\label{D_h(t)}
\Delta_h=\Delta_0\exp\Bigg(\int_1^{h/\Delta_h}\frac{e^{-\delta_\text{LZ}/s^2}-1}{\sqrt{s^2-1}}ds \Bigg),
\end{gather}
where $\delta_\text{LZ}=\pi\alpha^2k_F^2/\gamma$.
The numerically integrated function $\Delta_h[h(t)]$ is shown in Fig. \ref{fig4} and  below we discuss its behavior for different tunneling regimes.

(i) The value $\delta_\text{LZ}=0$ corresponds to zero RSOC effects ($\alpha=0$), so that the spectral branches do not change after crossing and the trivial solution for the gap $\Delta_h=\Delta_0$ holds.

(ii) The limit of $\delta_\text{LZ}\ll 1$ means that $\gamma\gtrsim \alpha^2k_F^2$ and the spectral branches intersect nonadiabatically, or so rapidly that they do not feel the RSOC. The Landau-Zener tunneling is suppressed and the gap has a weak dependence on the Zeeman field at $h(t)>\Delta_0$: 
\begin{gather} \notag
\Delta_h\approx \Delta_0\exp\Big(-\delta_\text{LZ}\frac{\sqrt{h^2(t)-\Delta_h^2}}{h(t)} \Big).
\end{gather}

(iii) In the opposite limit of $\delta_\text{LZ}\gg 1$  with $\gamma\ll \alpha^2k_F^2$ the QPs undergo strong spin-flip tunneling during an almost adiabatic avoided crossing. This leads to the effective formation of the triplet superconducting correlations (or related triplet component of the anomalous Green function \cite{gorkov_superconducting_2001}) even for the small RSOC energy $\alpha k_F/\Delta\ll 1$. Such dynamically generated correlations are determined by the rate of field change $\gamma$ and their effect on the gap can significantly exceed the static mixing of singlet-triplet pairs for $\alpha\neq 0$ \cite{Tewari_2011}.
As a result, the singlet gap function (\ref{I_SCE}) is suppressed and the self-consistency equation reads 
\begin{gather} \notag
\Delta_h\approx 
\begin{cases}
\sqrt{\Delta_0(2h(t)-\Delta_0)} 
\quad \text{for} \quad \Delta_h>h/\delta_\text{LZ}, \\ 
\Delta_0\exp\Big(-\delta_\text{LZ}\frac{\sqrt{h^2(t)-\Delta_h^2}}{h(t)} \Big)
\frac{\exp(\sqrt{\delta_\text{LZ}}\sqrt{\delta_\text{LZ}-1})}{\sqrt{\delta_\text{LZ}}+\sqrt{\delta_\text{LZ}-1}}
\\
\quad \text{for} \quad \Delta_h<h/\delta_\text{LZ}.
\end{cases}
\end{gather}
(iv) The critical value $\delta_\text{LZ}\rightarrow\infty$ corresponds to the complete Landau-Zener spin-flip  tunneling, so that there are no QPs at the energies $E>0$. 
In this case we have restored the thermodynamically metastable branch $\Delta_h\approx \sqrt{\Delta_0(2h(t)-\Delta_0)}$ from well-known static case \cite{SARMA19631029}.

The actual behavior of the gap  in time must be determined by switching between different branches of $\Delta_h[h]$ as the Zeeman field $h(t)$ increases. The first solution, which is fixed by the initial condition $\Delta_h(t=0)=\Delta_0$ holds until $h(t)=\Delta_0$, where $\Delta_h$ goes to another unique possible solution $\Delta_h[h]$ for a given $\delta_\text{LZ}$ (see the red point and black dashed line in Fig. \ref{fig4}). The question of the exact dynamics of the gap in the jump region is difficult, because due to the rapid change in the $\Delta_h$, the decoupling of the LZSM problem and self-consistency equation may not be guaranteed [Section \ref{sec_4_B}]. It is qualitatively expected that the jump at $t\approx\Delta_0/\gamma$ should be smeared both by non-zero static contribution of SOC to the gap (since the equilibrium gap value depends on $\alpha$) and by the QP tunneling dynamics. At large times $h(t)\gg\Delta_0$ there are no transitions between the QP states ($p_k\rightarrow 1$), since the splitting between the spectral branches becomes zero and therefore the gap tends to the constant asymptotics $\Delta_h(\infty)$. 
%

\subsection{QP interference effects}\label{sec_4_D}

In addition to the dominating term $\Delta_h$, the gap equation (\ref{D_h}) also contains small rapidly oscillating term  
\begin{gather}\label{dd_t}
\delta\Delta(t) 
=\lambda \sum_{|\xi_k|<\sqrt{h^2-\Delta^2}} 
    u_0u_1 ie^{-i\theta} C_{k\uparrow+}^*C_{k\downarrow-}
    \\ \notag
    +v_0v_1 (-i)e^{i\theta} C_{k\downarrow-}^*C_{k\uparrow+},
\end{gather}
arising from the interference of the QP states which have experienced LZSM transitions. 
It is obvious that in its structure this function resembles the collective Higgs mode, which is excited in a natural way during the redistribution of states in the QP spectrum. Let us look at it in more details. 
Using the time-dependent coefficients (\ref{C_0}-\ref{C_2}) we obtain
\begin{gather}\label{minor_gap}
\frac{\delta\Delta(t)}{\lambda N(0)}
= \int_{-\sqrt{h^2-\Delta^2}}^{\sqrt{h^2-\Delta^2}}  \sqrt{p_k}\sqrt{1-p_k} G(\xi, t) \cos(D_k(t)) d\xi,
\end{gather}
where we introduce the dynamical phase $D_k(t)=2 \int_{t_0}^t (E_0-h(t))dt+\chi_k+\pi$ and the function $G(\xi,t)=\text{sgn}(\alpha)(u_0u_1+v_0v_1)$. 
The function $G(\xi)$ is proportional to $\alpha k_F/\Delta$, which means that $\delta \Delta(t)$ is parametrically small and can be considered against the background of the main change in the gap $\Delta_h $ from the equation (\ref{D_h(t)}).

For the integral (\ref{minor_gap}), it is easy to estimate the asymptotic behavior at large times $\Delta_0/\gamma \ll t$. The dynamical phase is written as $D_k(t)=-(E_0^2+\gamma^2t^2)/\gamma+\chi_k+\pi + 2E_0t $ for the spin-splitting field $h(t)=\gamma t$. Here $2E_0t$ is a fast oscillating term at $t\rightarrow \infty$ and one can use a stationary phase approximation for the $\xi$-integration in Eq. (\ref{minor_gap}) with the stationary phase point $\xi=0$. Using Eq. (\ref{G_func}) for $G(\xi=0,t)$, we find the asymptotic behavior of $\delta\Delta(t)$:
\begin{gather}\label{ampl}
\delta\Delta(t) \approx \lambda N(0) e^{-\delta_\text{LZ}/2}\sqrt{1-e^{-\delta_\text{LZ}}} \frac{|\alpha| k_F}{2(\gamma t-\Delta_h)}\sqrt{\frac{\pi\Delta_h}{t}} 
\\ \notag \times
\cos\Big(\frac{(\gamma t-\Delta_h)^2}{\gamma} +\frac{3\pi}{4} -\chi_0 \Big), 
\end{gather}
where $\Delta_h[h(t\rightarrow \infty)]$ from Eq. (\ref{D_h(t)}) is a constant determined by $\delta_\text{LZ}$.
The result obtained means that the collective interference between the two QP states at each $\xi_k$ after LZSM crossing behaves at large times as a modified Higgs mode. Due to linear dependence $h(t)$, this mode has a modulated frequency and polynomial damping law $\propto t^{-3/2}$ arising from the inhomogeneous broadening of the mode. 
Note that for large times only the contribution from the point $\xi=0$ survives, so the amplitude of $\delta\Delta(t)$ does not depend on the number of redistributed states in the QP spectrum.

If the linear growth of the spin-splitting field $h(t)$ stops at a certain value $h_f>\Delta_0$ after the redistribution of some of the QP states, then the accumulated dynamic phase $D_k(t)$ and the gap fluctuation will depend only on this value $h_f$
\begin{gather}
\delta\Delta(t) \approx 
\lambda N(0) e^{-\delta_\text{LZ}/2}\sqrt{1-e^{-\delta_\text{LZ}}} \frac{|\alpha| k_F}{2(h_f-\Delta_h)} \\ \notag
\times\sqrt{\frac{\pi\Delta_h}{t}} \cos\Big(2(h_f-\Delta_h) t-\frac{3\pi}{4}
-\frac{\Delta_h^2-h_f^2}{\gamma}  +\chi_0 \Big) .
\end{gather}
The specific spectral distortion occurring between two brancher $E_{k\uparrow+}$ and $E_{k\downarrow-}$ during the Landau-Zener dynamics at $h(t)<h_f$ acts as an initial perturbation for the gap function at $t=h_f/\gamma$. The free gap dynamics at $t>h_f/\gamma$ resembles the Higgs mode with $\delta\Delta(t) \propto  \cos(2(h_f-\Delta_h)t)/\sqrt{t} $ at the frequency $\omega=2|\Delta_h-h_f|$ (or $\omega=\omega_-$ in our previous notations) with the standard damping law. It is interesting, that the amplitude of this mode proportional to $|\alpha| k_F$ instead of $\alpha^2k_F^2$ as it is expected in the case of small perturbations [Section (\ref{sec_3})]. Such amplification is a direct consequence of the intersection of two specific spectral branches and the subsequent non-adiabatic dynamics. Thus, this mode turns out to be leading in comparison with other nonadiabatic corrections arising due to the interaction of all QP spectral branches. Note, that the method for calculating the self-consistency equation developed in Section \ref{sec_3} can be combined with the Landau-Zener problem (\ref{LZ_full}) and all corrections can be computed within the perturbation theory.

\subsection{Density of states and distribution function}\label{sec_4_E}

Rearrangment of the spectrum as a result of the intersection of spectral branches naturally leads to a change of the structure of the density of states (DOS), that has become time dependent. Since the temporal evolution of the spectrum is adiabatic except the small region where the crossing occurs one can use the quasistatic description of the DOS. 
For the small RSOC the DOS for one spin projection can be written in terms of Bogoliubov-de Gennes functions
\begin{gather}
N_\uparrow(E,t)\approx \sum_{k}\sum_{n=\uparrow+,\uparrow-} |u_0|^2\delta\Big(E-E_{kn}[h(t)]\Big) \\ \notag
+|v_0|^2\delta\Big(E+E_{kn}[h(t)]\Big).
\end{gather}
Here we use static QP amplitudes $u_0$ and $v_0$ (see Eq. \ref{bdgf}) to distinguish the particle/hole contributions and $E_{k\uparrow\pm}$ are defined in Eq. (\ref{spectrum}). 
The calculation of $N_\uparrow$ is cumbersome, because the RSOC shifts the spectral branches and opens a minigap $\propto \alpha k_F$ at $E=0$ [Appendix \ref{APP_E}]. For the small RSOC parameter these changes are negligible and one can use a standard expression for the DOS, which now depends on time through the spin-splitting field 
\begin{gather}\label{DOS_}
\frac{N_\uparrow(E,t)}{N(0)}\approx 
\frac{|E+h(t)|}{\sqrt{(E+h(t))^2-\Delta_h[h(t)]^2}}.
\end{gather}
Here the gap function $\Delta_h$ is taken from (\ref{D_h(t)}) and two coherence peaks are present at $E = \pm \Delta_h[h(t)]-h(t)$.

The amplitude of the QP wavefunction $\psi_k(t)$ from (\ref{wf}) contains the information about filling (or occupation) of the $\xi_k-$th state. More precisely the coefficients $|C_{k\uparrow\pm}(t)|^2$ and $|C_{k\downarrow\pm}(t)|^2$ can serve as an effective distribution functions $f_{\uparrow\downarrow}(E)$ for QPs with different spin projections. 
As it was discussed in the section \ref{sec_4_C}, the temporal evolution of these coefficients is defined by the LZSM problem, and for spin-up states one has 
\begin{align}\notag
    |C_{k\uparrow-}(t)|^2 &= 1, \\ \notag
    |C_{k\uparrow+}(t)|^2 &= (1-p_k) \Theta\big[ \sqrt{h^2(t)-\Delta_h^2} - \xi_k \big],
\end{align}
which can be rewritten as a distribution function
\begin{gather}\label{distrib}
f_{\uparrow}(E, t) \approx 
\begin{cases}
    0, \quad E>0 \\ 
    1-\exp\Big[-\frac{\delta_\text{LZ}\Delta_h^2[h(t)]}{(E+h(t))^2} \Big], \quad \Delta_h-h< E<0 \\
    1, \quad E<\Delta_h-h
\end{cases}
\end{gather}
The dependence $f_\uparrow(E)$ is shown in Fig. \ref{fig4} for $\delta_\text{LZ}=0.5$.
The most pronounced change of the distribution function occurs at $E\approx \Delta_h-h$, since for large QP energies the LZSM tunneling is suppressed.
For the opposite spin projection the DOS $N_{\downarrow}(E)$ has the similar structure (\ref{DOS_}) with $h\rightarrow -h$, while the corresponfing distribution function $f_{\downarrow}(E)$ is different and is given by the Eqs. (\ref{C_0}-\ref{C_2}). The DOS structure and effective distribution function enable the calculation of a system's optical or transport response, which can be experimentally measured.

\subsection{Dynamical magnetization of QP gas }
\label{sec_4_F}

Nonadiabatic LZSM tunneling of QP states causes a spin imbalance in the spectrum, which results in the appearance of nonzero dynamical magnetization. Using the notations from the previous section we get an expression for the z-component of the magnetization per unit volume
\begin{gather}
m_z(t)=
\mu_B \sum_\text{i.c.} 
\check\psi^\dagger_{k}(t) \check{\tau}_m \check\psi_{k}(t),
\end{gather}
where $\check{\tau}_m=(\hat{\tau}_0+\hat{\tau}_z)\otimes \hat{\sigma}_z/2$; the vector  $\check\psi_{k}(t)$ is a solution of the TDBdG problem (\ref{wf}) and "i.c." here means the summation over all initial conditions (see Eq. (\ref{I_SCE})).
Due to symmetry and homogenuty of the problem for the field ${\bf h}(t)= h(t){\bf z}_0$ the transversal components of the magnetization $m_{x,y}(t)$ are zero.

\begin{figure}[] 
\centering
\begin{minipage}[h]{1.0\linewidth}
\includegraphics[width=1.0\textwidth]{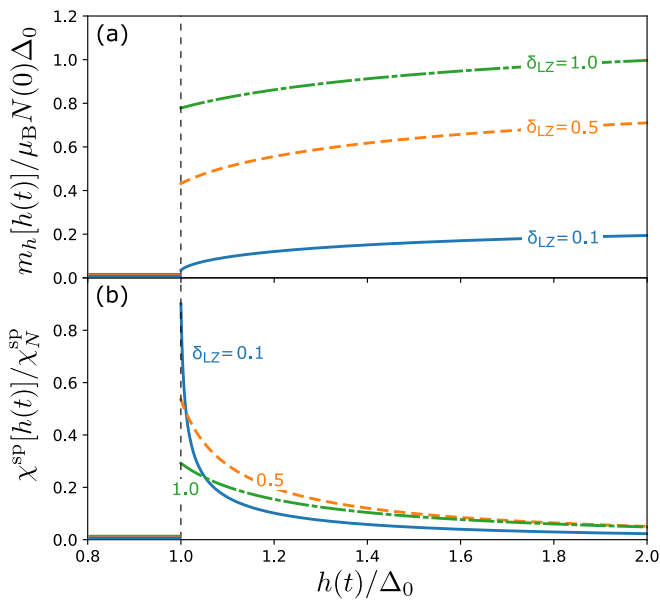} 
\end{minipage}
\caption{ {\small 
(a) Dynamical magnetization $m_h$ per unit volume induced by the nonadiabatic tunneling of QP states and (b) corresponding spin susceptibility $\chi^\text{sp}$  versus time-dependent spin-splitting field $h(t)$ for different values of $\delta_\text{LZ}$.  }}
\label{fig5}
\end{figure}

Taking the dynamical amplitudes $C_n(t)$ from (\ref{C_0}-\ref{C_2}) and implementing the same procedure as for the self-consistency equation (\ref{D_h}-\ref{d_h+dd}) we found that the magnetization can be written as follows
\begin{gather}\label{m_z}
m_z(t) = m_h[h(t)]+\delta m(t) .
\end{gather}
As in the case of the gap equation (\ref{d_h+dd}) we have two contributions: $m_h[h(t)]$ which is a slow function of time arising from the redistribution of the quaiparticle states, and $\delta m(t)\propto \alpha k_F$ which is small oscillatory term originated from the interference of the redistributed states. The first term can be easily calculated with the help of the quasiparticle density
\begin{gather}
    n_{\sigma}(t)=\int N_{\sigma}(E,t)f_{\sigma}(E,t) dE,
\end{gather}
where $\sigma=\uparrow\downarrow$ and the DOS $N_\sigma(E)$ and distribution function $f_\sigma(E)$ are defined in the previous subsection. Corresponding spin imbalance results in the dynamical magnetization $m_h[h(t)]=\mu_B(n_\uparrow-n_\downarrow)$, which is shown in Fig. \ref{fig5}(a).

For $h(t)<\Delta_0$ there is no crossing of the QP spectral branches and according to our model there is no tunneling between QP states, therefore the dynamical magnetisation is zero. Once the intersection has occured at $h(t)=\Delta_0$, the distribution functions $f_{\uparrow,\downarrow}(E)$ transform and nonzero spin imbalance $n_\uparrow-n_\downarrow$ is generated. Due to the jump of $\Delta_h$ function at $h(t)=\Delta_0$ [Fig. \ref{fig4}] the magnetization at this point also has a sharp discontinuity. 
At large times the tunneling of QP states is suppressed therefore the magnetization is saturated to a constant value determined by the parameter $\delta_\text{LZ}$.  
Obviously, an increase in $\delta_\text{LZ}$ makes the spin-flip tunneling more efficient and thereby increases the maximum value of $m_h$.
The second therm in (\ref{m_z}) resembles the Higgs mode term (\ref{dd_t}) and gives negligible contribution to $m_z(t)$, therefore it can be discarded.

In addition one can compute the dynamical susceptibility of the QP gas in the Zeeman field of the general form $h(t)=\mu_\text{B}H(t)$. 
It is known that an orbital and a spin parts of the magnetic susceptibility can be splitted in the case of small spin-orbital effects \cite{gorkov_superconducting_2001}. Since we consider a homogeneous system and neglect all orbitals effects only the spin part plays a role, which can be written as follows
\begin{gather}\label{chi_sp}
\chi^{\text{sp}}[h(t)] = 
\mu_B\frac{\partial m_h}{\partial h}.
\end{gather}
The ratio of the numerically calculated susceptibility $\chi^{\text{sp}}[h(t)]$ and the normal susceptibility $\chi_{N}^\text{sp}=2\mu_B^2N(0)$ \cite{Frigeri_2004} is shown if Fig. \ref{fig5}(b). It is seen that spin-flip tunneling in the QP spectrum provokes a paramagnetic response of the superconducting condensate. The function (\ref{chi_sp}) should have a singularity $\chi^{\text{sp}}\propto (h(t)-\Delta_0)^{-1/2}$ in the vicinity of $h(t)\approx\Delta_0$, which is defined by the shape of the QP spectrum at $k\approx k_F$ and has the same origin as the coherence peak in the DOS (\ref{DOS_}). However, due to the jump of the order parameter $\Delta_h$ at this point we observe shifted peaks, which have to be smeared out near $h(t)=\Delta_0$ if more realistic model of LZSM tunneling [Section \ref{sec_4_B}] is taken into account. We note again that we discuss only the dynamic contribution to the susceptibility, which, generally speaking, have to be added to the static one, which is not equal to zero at $T=0$ in the presence of SOC \cite{gorkov_superconducting_2001, Frigeri_2004}.

\section{Discussion and experimental perspectives}\label{sec_5}

We analyzed the coherent dynamics of the superconducting condensate in the presence of Zeeman field and SOC in collisionless regime. First, it was established that the Higgs mode of the superconducting gap is sensitive to the spin-splitting field $h_0$ and can be directly triggered by either its harmonic perturbation $\delta h(t)$ or by an external laser pulse. Second, it was shown, that the field $h(t)\sim t$ can provoke an avoided crossing of the QP spectral branches and adiabatic spin-flip tunneling of the QPs between the different branches occurs. Corresponding redistribution of the QPs in the spectrum leads to the appearance of the dependence $\Delta[h(t)]$ and generation of the interference effects. Emerging spin imbalance reveals itself in the effective dynamical distribution function and in a generation of a weak magnetization of the QP gas. 

We propose superconductor-ferromagnet hybrid structures as an experimental platform for detecting the described effects. The ferromagnetic layer can serve as a source of both Rashba spin-orbit coupling and an exchange field. Since it is important to remove orbital effects from the system, the most suitable geometry for superconductor is either thin film or one-dimensional nanowire \cite{szombati_josephson_2016}. 

The excitation and observation of Higgs modes in superconductors requires frequencies of the order of $\Delta_0/\hbar$, which vary from the far infrared to the terahertz range. The laser excitation of modes seems to be the most practical and feasible, and the detection can be implemented using the THz light source with ultrafast pump-probe spectroscopy or third harmonic generation measurements \cite{PhysRevB.76.224522, PhysRevLett.111.057002, doi:10.1126/science.1254697}. Generation of a fast oscillating component of the homogeneous Zeeman field $\delta h(t)$ inside a superconductor is a difficult task especially for the THz range. Some proposals can be made amid encouraging progress in the ultrafast optical control of magnetization in various materials \cite{kirilyuk_ultrafast_2010, kirilyuk_laser-induced_2013,ELGHAZALY2020166478}. Ferromagnetic resonance induces the time-dependent stray field which in combination with geometric constraints may serve as a Zeeman field inside a thin superconducting film, as it was discussed for a two-dimensional electron gas in Ref. \cite{plekhanov_floquet_2019}. Upon the excitation of the Higgs modes by the Zeeman field the THz spectroscopy measurements can be implemented again.

It is possible to make basic parameter estimates for the experimental observation of LZSM transitions in the QP spectrum. For example, consider $\Delta_0=0.1$meV (for $T_c\approx 1$K) and $\alpha k_F \sim 10^{-3}$meV $\ll \Delta_0$. Then the constraint for the small tunneling rate is $\hbar \gamma \gtrsim \alpha^2 k_F^2$ in dimensional units, which is equivalent to $\gamma \gtrsim  10^{-3}$meV/ns. Consider inelastic relaxation of QP with a typical time $\tau_\text{ph}\sim 100$ns in the case of the electron-phonon scattering at low temperatures \cite{gershenzon1990electron,kardakova_electron-phonon_2013} . The collisionless regime is maintained at $t\ll \tau_\text{ph}$, which corresponds to times $t\lesssim 10$ns. 
Under such conditions the field $h(t)=\gamma t \sim \Delta_0$ is achievable only for $\gamma \sim 10^{-2}$meV/ns. The measurements of the various properties of the superconducting condensate above Pauli limit at short times can be implemented with the help of ultrafast THz techniques, such as pump-probe \cite{PhysRevLett.111.057002} for an optical conductivity. 


\acknowledgements
This work has been supported by ANR OPTOFLUXONICS, ANR SUPERFAST, the LIGHT S$\&$T Graduate Program and the Russian Science Foundation (Grant No. 21-72-10161). A.S.M. acknowledges support from the State Contract of Ministry of Science and Higher Education of Russian Federation No. 075-03-2022-106 (project FSMG-2023-0011) of Moscow Institute of Physics and Technology.

\appendix

\section{Eigenvectors }\label{APP_A}

The instantaneous eigenvectors of the Hamiltonian $\check{\mathcal{H}}(k,t)$ from Eq. (\ref{TDHam}) can be written as follows
\begin{gather}\label{A_1} 
\check{\Psi}_{kn}(t)=\frac{1}{\sqrt{1+a_{1n}^{2}+a_{2n}^2+a_{3n}^2}}
\begin{pmatrix}
    1 \\
    -ia_{1n}e^{i\theta_k} \\
    -ia_{2n}e^{i\theta_k} \\
    a_{3n} 
\end{pmatrix},
\end{gather}
where we have defined the phase $\theta_k=\arg\big(k_x+ik_y\big)$ and real coefficients
\begin{gather}\label{A_2}
a_{1n}=\frac{(h+E_{kn})^2-E_0^2-\alpha^2k^2}{2\alpha k(\xi_k+h)}, \\ \notag
a_{2n}=\frac{\alpha k}{\Delta}-\frac{E_{kn}-\xi_k-h}{\Delta}a_{1n}, \\ \notag
a_{3n}=\frac{E_{kn}-\xi_k+h}{\Delta}-\frac{\alpha k}{\Delta}a_{1n}.
\end{gather}
The instantaneous eigenvalues of $\check{\mathcal{H}}(k,t)$ are 
\begin{gather} \notag
    E_{kn}(t) \equiv E_{k\sigma\pm}(t) = \\ \notag
    \pm \sqrt{ E_0^2+\alpha^2k^2+h^2(t) \mp \text{sgn}(\sigma) 2\sqrt{\xi_k^2\alpha^2k^2+h^2(t)E_0^2} },
\end{gather}
where $E_0=\sqrt{\xi_k^2+\Delta^2}$; the subscript $\pm$ refers to spectral branch above/below the Fermi level and $\sigma=\{\uparrow, \downarrow\}$ denotes a spin subband. 
Note, that the Hamiltonian (\ref{TDHam}) implies the symmetry relations between the energies $E_{k \uparrow+}= -E_{k \downarrow-}$ and $E_{k\downarrow +}= -E_{k \uparrow-}$, and between the corresponding eigenvectors $\check{\Psi}_{k \uparrow+}= i\hat{\tau}_y\otimes\hat{\sigma}_z \check{\Psi}_{k \downarrow-}^{*}$ and $\check{\Psi}_{k\downarrow +}= i\hat{\tau}_y\otimes\hat{\sigma}_z \check{\Psi}_{k \uparrow-}^{*}$, where $\hat{\tau}_i(\hat{\sigma}_i)$ is the Pauli matrix in the Nambu(spin) space.

For the case of weak SOC  $\alpha k_F\ll \{E_F, h(t), \Delta(t)\}$ the eigenvectors (\ref{A_1}) can be expanded up to the first order in $\alpha k_F/\Delta$ as follows 
\begin{gather} \label{APP_A_eigenf}
\check{\Psi}_{k\uparrow+}
\approx
\begin{pmatrix}
    u_{0} \\
    -iu_{1}e^{i\theta_k} \\
    -iv_{1}e^{i\theta_k} \\
    v_{0}   
\end{pmatrix},
\quad
\check{\Psi}_{k\downarrow+}
\approx
\begin{pmatrix}
    iu_{1}e^{-i\theta_k}\\
    -u_{0} \\
    v_{0}  \\
    -iv_{1}e^{-i\theta_k}
\end{pmatrix}, 
\\ \notag
\check{\Psi}_{k\uparrow-}
\approx
\begin{pmatrix}
    -v_0 \\
    iv_{1}e^{i\theta_k} \\
    -iu_{1}e^{i\theta_k}  \\
    u_0 \\
\end{pmatrix},
\quad
\check{\Psi}_{k\downarrow-}
\approx
\begin{pmatrix}
    -iv_{1}e^{-i\theta_k} \\
    v_{0} \\
    u_{0} \\
    -iu_{1}e^{-i\theta_k} \\  
\end{pmatrix}.
\end{gather}
Here we define equilibrium QP amplitudes
\begin{gather} \label{bdgf}
    u_0=\frac{1}{\sqrt2}\sqrt{1+\frac{\xi_k}{\sqrt{\xi_k^2+\Delta^2}}},  
    \quad
    v_0=\frac{1}{\sqrt2}\sqrt{1-\frac{\xi_k}{\sqrt{\xi_k^2+\Delta^2}}},
\end{gather}
and $u_1, v_1 \propto \mathcal{O} (\alpha k_F/\Delta)$ correspond to the triplet component of the QP wavefunctions. 

The function $G(\xi,t)=\text{sgn}(\alpha)(u_0u_1+v_0v_1)$ from Eq. (\ref{minor_gap}) can be found at the stationary phase point $\xi=0$ by expanding the coefficients (\ref{A_2}) in the small parameter $\alpha k_F/\Delta$ and putting $\Delta\approx \Delta_h$ (see Eq. (\ref{D_h(t)})). The result reads 
\begin{gather}\label{G_func}
G(0,t) \approx \frac{|\alpha| k_F}{2(h(t)-\Delta_h)}.
\end{gather}

\section{Derivation of linearized self-consistency equation}\label{APP_B}

We start with the linearized (\ref{III_dCdt}, \ref{linearization}) dynamical equations
\begin{gather}\label{eqC}
i\frac{\partial}{\partial t}\delta C_{km}(t)=
\sum_n  \check{\Psi}_m^{\dagger}\check{\mathcal{V}}(t)\check{\Psi}_n e^{- i (E_n-E_m) t}\big( \delta_{n,l}+\delta C_{kn}(t) \big),
\end{gather}
where the indices $n,m=\{\uparrow+, \downarrow+, \uparrow-, \downarrow-\}$ and two possible initial configurations (\ref{i_c_}) are marked as ${l=\{\uparrow-, \downarrow-\}}$.
The compact form of the self-consistency equation for the gap (\ref{III_gap_full}) is
\begin{gather}\label{B_SCE}
\Delta_\text{eq} + \delta\Delta(t) = \\ \notag
-\frac{\lambda}{2}\sum_l\sum_k\sum_{n,n'}
\big( \delta_{n,l}+\delta C_{kn}(t)^* \big)
\big( \delta_{n',l}+\delta C_{kn'}(t) \big) \\ \notag
\times
e^{-i(E_{n'}-E_n)t}\check{\Psi}_{kn}^{\dagger} \check{\tau}_{\Delta}  \check{\Psi}_{kn'}.
\end{gather}
As was mentioned in Section \ref{sec_2}, we neglect the effect of RSOC on the equilibrium value of the gap, which can be taken as $\Delta_\text{eq}=\Delta_0$. It also makes sense to omit the negligibly small corrections from the RSOC to the energy spectrum, so one can put $E_n\equiv E_{k\sigma \pm} \approx\pm  E_0- \text{sgn}(\sigma)h_0$.

The equations (\ref{eqC}, \ref{B_SCE}) can be simplified and written as follows 
\begin{gather}\label{whole_system}
\frac{\partial f_1}{\partial t} = i e^{i(2(E_0-h_0)t)}[\mathcal{A}\delta \Delta(t)-\mathcal{B}\delta h(t)], \\ \notag
\frac{\partial f_2}{\partial t} = i e^{i(2(E_0+h_0)t)}[\mathcal{A}\delta \Delta(t)+\mathcal{B}\delta h(t)],  \\ \notag
\frac{\partial g}{\partial t} = i \frac{\xi}{E_0}e^{i(2E_0t)}\delta \Delta(t), \\ \notag
    \delta\Delta(t)= 
    \Big\langle  \frac{\mathcal{A}}{2} \text{Re}f_1(t)e^{-i(2(E_0-h_0)t)} \Big\rangle \\ \notag
    + \Big\langle  \frac{\mathcal{A}}{2} \text{Re}f_2(t)e^{-i(2(E_0+h_0)t)} \Big\rangle 
    + \Big\langle  \frac{\xi}{E_0} \text{Re} g(t) e^{-i(2E_0t)} \Big\rangle.
\end{gather}
We have used the notation 
$\big\langle \dots \big\rangle = \lambda\sum_k \approx \lambda N(0) \int_{-\omega_D}^{\omega_D}d\xi$
and introduced new complex-valued functions
\begin{gather}
f_1\equiv -ie^{i\theta}\delta C_{\uparrow+}, \quad g_1\equiv -\delta C_{\downarrow+}, \\ \notag
f_2 \equiv -ie^{-i\theta}\delta C_{\downarrow+}, \quad g_2\equiv -\delta C_{\uparrow+}, \\ \notag
g = \frac{g_1+g_2}{2}, 
\end{gather}
where the subscript corresponds to the two possible initial conditions. 
The functions 
\begin{align}\label{AandB}
\mathcal{A}(\xi)&=2(u_0u_1+v_0v_1) \propto \mathcal{O} (\alpha k_F/\Delta_0), \\ \notag
\mathcal{B}(\xi)&=2(u_0v_1+u_1v_0) \propto \mathcal{O} (\alpha k_F/\Delta_0)
\end{align}
have the lowest order in $\alpha k_F$ parameter [Appendix \ref{APP_A}] and are even in $\xi$.
All terms odd in $\xi$ in Eq. (\ref{whole_system}) are related to the imaginary part of $\delta\Delta(t)$ and vanish due to the approximate electron-hole symmetry of BdG Hamiltonian (\ref{TDHam}), due to which the density of states is approximated as $N(\xi)\approx N(0)$ in the $\langle ... \rangle$-integration \cite{tsuchiya_hidden_2018}.

Applying the Laplace transform $ f(s)=\int_0^{\infty}e^{-st} f(t) dt$ with $s=i\omega+\zeta$  (where $\zeta \rightarrow 0$) for Eq. (\ref{whole_system})  we obtain the gap equation in the complex plane, which is found to be
\begin{widetext}
\begin{gather}\label{B_SCE2}
    \delta\Delta(s) 
    = \delta\Delta(s) \Big\langle \frac{2\xi^2}{E_0}\frac{1}{s^2+4E^2_0} \Big\rangle
    + \delta\Delta(s) \Big\langle \mathcal{A}^2(\xi)\frac{(E_0+h_0)}{s^2+4(E_0+h_0)^2} \Big\rangle 
    + \delta\Delta(s) \Big\langle \mathcal{A}^2(\xi)\frac{(E_0-h_0)}{s^2+4(E_0-h_0)^2} \Big\rangle
    \\ \notag
    + \delta h(s) \Big\langle \mathcal{A}(\xi)\mathcal{B}(\xi)\frac{(E_0+h_0)}{s^2+4(E_0+h_0)^2} \Big\rangle
    - \delta h(s) \Big\langle \mathcal{A}(\xi)\mathcal{B}(\xi)\frac{(E_0-h_0)}{s^2+4(E_0-h_0)^2} \Big\rangle
    \\ \notag
    +\Big\langle f_1'(0) \frac{\mathcal{A}(\xi)}{2} \frac{s}{s^2+4(E_0-h_0)^2} \Big\rangle 
    + \Big\langle f_1''(0) \frac{\mathcal{A}(\xi)(E_0-h_0)}{s^2+4(E_0-h_0)^2} \Big\rangle 
    \\ \notag
    + \Big\langle f_2'(0) \frac{\mathcal{A}(\xi)}{2} \frac{s }{s^2+4(E_0+h_0)^2} \Big\rangle 
    + \Big\langle f_2''(0) \frac{\mathcal{A}(\xi)(E_0+h_0)}{s^2+4(E_0+h_0)^2} \Big\rangle \\ \notag
    + \Big\langle g'(0)  \frac{\xi}{E_0}\frac{s }{s^2+4E^2_0} \Big\rangle 
    + \Big\langle g''(0) \frac{2\xi E_0}{s^2+4E^2_0} \Big\rangle.
\end{gather}
\end{widetext}
Here $f=f'+if''$ and the initial conditions $f_{1,2}(0)=f_{1,2}(t=0)$, $g(0)=g(t=0)$ implicitly contain the initial value of the gap perturbation $\delta \Delta(t=0)$. Note that terms with initial conditions will be discarded when calculating the superconductor response (see Section \ref{sec_3}).
Now we can single out functions of $s$ with different singularities in the complex plane and denote them using short notations
\begin{gather}\label{new_func}
    \mathcal{K}_0(s) = \Big\langle \frac{2\xi^2}{E_0}\frac{1}{s^2+4E^2_0} \Big\rangle,  \\ \notag
    \mathcal{K}_\pm(s) = \Big\langle \mathcal{A}^2(\xi)\frac{(E_0\pm h_0)}{s^2+4(E_0\pm h_0)^2} \Big\rangle,  \\ \notag
    \mathcal{F}_\pm(s) = \Big\langle \mathcal{A}(\xi)\mathcal{B}(\xi)\frac{(E_0\pm h_0)}{s^2+4(E_0\pm h_0)^2} \Big\rangle
\end{gather}
and $\mathcal{I}(s)$, which consists of all initial perturbations at $t=0$. 
The functions $\mathcal{A}$ and $\mathcal{B}$ are of the first order in the small parameter $\alpha k_F/\Delta$, therefore we have
\begin{gather*}
\mathcal{K}_0(s) \propto \mathcal{O}\Big(\frac{\alpha^0 k^0_F}{\Delta_0^0}\Big), \\
\mathcal{K}_\pm(s), \mathcal{F}_\pm(s) \propto \mathcal{O}\Big(\frac{\alpha^2 k^2_F}{\Delta_0^2}\Big).
\end{gather*}

It can be shown that the the difference $[\mathcal{F}_+(s)-\mathcal{F}_-(s)]$ is proportional to $h_0$. This allows one to write the terms with $\delta h(t)$ in (\ref{B_SCE2}) as 
$\delta h(s)[\mathcal{F}_+(s) - \mathcal{F}_-(s)]$
or
$({\bf h}_0 \cdot \delta {\bf h}(s)) [\mathcal{F}_+(s) - \mathcal{F}_-(s)]/h_0$,
where both vectors are oriented along ${\bf z}_0$ axis.
By rewriting the equation (\ref{B_SCE2}) with the new introduced functions (\ref{new_func}) we get the self-consistency equation (\ref{III_gap}).

\section{Long-time behavior of $\delta\Delta(t)$}\label{APP_C}

The susceptibility $\text{Im}\chi_{\Delta h}(s)|_{\zeta\rightarrow 0}=\text{Im}\chi_{\Delta h}(\omega)$ in Eq. (\ref{III_integral}) has  strongly dominant terms in the vicinity of different branch points in the interval $\omega\in[\omega_-, \infty)$. In order to demonstrate this, the function $\text{Im}\chi_{\Delta h}(s)$ can be expanded in a series up to the second order in the parameter $\alpha k_F/\Delta$, and this expansion must be carried out accurately near the branch points and may differ in different regions of $\omega$.
Therefore, we assume that the value of the integral is determined by these dominant contributions of $\text{Im}\chi_{\Delta h}(\omega)$ and can be evaluated sequentially as
$\int_{\omega_-}^{\infty} = \int_{\omega_-}^{\omega_0} + \int_{\omega_0}^{\omega_+} + \int_{\omega_+}^{\infty}.  $ 
Let us consider the small regions $ \Omega \ll \omega_{0,\pm}$ in the vicinity of these points separately. 

\begin{center}
{\bf 1:} $\omega\approx\omega_- + \Omega$
\end{center}
Close to the point $\omega=\omega_-$ the term $\mathcal{F}_-''(\omega)$ dominates:
\begin{gather}
\mathcal{F}_-''(\Omega) \approx -\lambda N(0)\frac{\pi \Delta_0 \mathcal{A}(0)\mathcal{B}(0) }{4  \sqrt{2\Delta_0 \Omega}} \propto \frac{1}{\sqrt{\Omega}}.
\end{gather}
Despite the kernel $1-\mathcal{K}_0'(\omega)$ goes to zero at $\omega\rightarrow \omega_0$ there is no singularity in $\chi_{\Delta h}(\omega)$ at this point due to the small terms of the order of $(\alpha k_F)^2$ in the denominator. Therefore, the region in the vicinity of $\omega_0$ will not contrubute to the intergal.
Thus, the behavior of the first integral for $\omega\in[\omega_-, \omega_0)$ at large time $h_0 t \gg 1$ can be estimated as follows
\begin{gather}\label{c_1}
\int_{\omega_-}^{\omega_0} \approx
 \text{Im} \Bigg[  \frac{\delta h(i\omega_-) e^{i\omega_- t}}{ \big[1- \mathcal{K}_0'(\omega_-) \big] }
\int_{0}^{\omega_0-\omega_-}  \mathcal{F}_-''(\Omega)
e^{i\Omega t} d\Omega  \Bigg] \\ \notag 
\approx
-\lambda N(0) \frac{\pi^{3/2} \Delta_0 \mathcal{A}(0)\mathcal{B}(0) }{4 \sqrt{2\Delta_0 t}} 
  \frac{\text{Im} \Big[  \delta h(i\omega_-) e^{i(\omega_- t + \pi/4)}   \Big]}{ \big[1- \mathcal{K}_0'(\omega_-) \big] }.
\end{gather}

\begin{center}
{\bf 2:} $\omega\approx\omega_0 + \Omega$
\end{center}
In the vicinity of the branch point $\omega=\omega_0$ the main contribution is defined by
\begin{gather}
\mathcal{K}_0''(\Omega) \approx -\lambda N(0) \frac{\pi }{2\Delta_0} \sqrt{\Delta_0 \Omega} \propto \sqrt{\Omega} .
\end{gather}
Thus at large times $h_0 t \gg 1$ we get
\begin{gather}\label{c_2}
    \int_{\omega_0}^{\omega_+} =
    \int_{\omega_0}^{\omega_+}  \frac{ \big[\mathcal{F}_+'(\omega)-\mathcal{F}_-'(\omega)\big]}{  \mathcal{K}_0''(\omega)   }
 \text{Im}\Big[ e^{i\omega t}  \delta h(i\omega) \Big]
d\omega 
\\ \notag
 \approx - \frac{2\sqrt{\Delta_0} }{ \sqrt{\pi t}} \frac{\big[\mathcal{F}_+'(\omega_0)-\mathcal{F}_-'(\omega_0)\big]}{\lambda N(0)}
  \text{Im} \Big[  \delta h(i\omega_0) e^{i(\omega_0 t + \pi/4)}   \Big].
\end{gather}

\begin{center}
{\bf 3:} $\omega\approx\omega_+ + \Omega$
\end{center}
For the last branch point $\omega = \omega_+$ the kernel $\mathcal{F}_+''(\omega)$ dominates:
\begin{gather}
\mathcal{F}_+''(\Omega) \approx \lambda N(0)\frac{\pi \Delta_0 \mathcal{A}(0)\mathcal{B}(0) }{4 h_0 \sqrt{2\Delta_0 \Omega}} \propto \frac{1}{\sqrt{\Omega}}. 
\end{gather}
At large times $h_0 t \gg 1$ we get
\begin{gather}\label{c_3}
    \int_{\omega_+}^{\infty} \approx
    \int_{\omega_+}^{\infty}  -\frac{ \big[1- \mathcal{K}_0'(\omega_+) \big]\mathcal{F}_+''(\omega)}{ \big[1- \mathcal{K}_0'(\omega_+) \big]^2 + \big[ \mathcal{K}_0''(\omega_+) \big]^2  } \\ \notag
    \times
 \text{Im}\Big[ e^{i\omega t} \delta h(i\omega) \big) \Big]
d\omega 
\\ \notag
\approx -\lambda N(0) \frac{\pi^{3/2} \Delta_0 \mathcal{A}(0)\mathcal{B}(0) }{4 \sqrt{2\Delta_0 t}} 
  \frac{\big[1- \mathcal{K}_0'(\omega_+) \big]}{ \big[1- \mathcal{K}_0'(\omega_+) \big]^2 + \big[ \mathcal{K}_0''(\omega_+) \big]^2 } 
  \\ \notag
\times
  \text{Im} \Big[ \delta h(i\omega_+) \big) e^{i(\omega_+ t + \pi/4)}   \Big].
\end{gather}

\begin{center}
{\bf 4:} total integral 
\end{center}
By combining all three contribution (\ref{c_1},\ref{c_2},\ref{c_3}) we will get the equation (\ref{dD}) in the main text. Note that discussed approximations work for $0<h_0<\Delta_0$. 
The functions $\mathcal{A}, \mathcal{B}$ from (\ref{AandB}) at the point $\xi=0$ can be calculated using the wavefunctions (\ref{APP_A_eigenf}). By expanding the coefficients (\ref{A_2}) we obtain
\begin{gather}
\mathcal{A}(0)\mathcal{B}(0) = \mathcal{A}^2(0) \approx \frac{(\alpha k_F)^2}{(\Delta_0-h_0)^2}.
\end{gather}
Also, the analytical expressions for the kernel $\mathcal{K}_0(\omega)$ at $\omega>0$ can be found:

\begin{gather}
\frac{1-\mathcal{K}_0'(\omega)}{\lambda  N(0)} =  \\ \notag
\begin{cases}
\frac{\sqrt{4\Delta_0^2-\omega^2}}{\omega} \arctan\Big( \frac{\omega} {\sqrt{4\Delta_0^2-\omega^2}}\Big) \quad  \text{for} \quad \omega < 2\Delta_0 \\
    -\frac{\sqrt{\omega^2-4\Delta_0^2}}{\omega}\frac12 \ln\Big( \frac{\omega-\sqrt{\omega^2-4\Delta_0^2}} {\omega+\sqrt{\omega^2-4\Delta_0^2}}\Big) \quad  \text{for} \quad \omega > 2\Delta_0
\end{cases},
\\ \notag
\ 
\\ 
\frac{\mathcal{K}_0''(\omega)}{\lambda N(0)} = -\frac{\pi}{2}\frac{\sqrt{\omega^2-4\Delta_0^2}}{\omega} \Theta[ \omega - 2\Delta_0 ].
\end{gather}
Finally, the expression with the kernels $\mathcal{F}_\pm(\omega)$ from (\ref{c_2}) can be calculated numerically for small $\alpha k_F \ll \Delta_0$:
\begin{gather}
   \frac{ \big[\mathcal{F}_+'(\omega_0)-\mathcal{F}_-'(\omega_0)\big]}{\lambda N(0)} \\ \notag
   = h_0\fint_{0}^{\omega_D} \mathcal{A}(\xi)\mathcal{B}(\xi) \frac{h_0^2 - \xi^2 - 2\Delta_0^2}{(\xi^2 - h_0^2)^2 - 4\Delta_0^2h_0^2} d\xi .
\end{gather}
%

\section{Derivation and solution of LZSM problem}\label{APP_D}

The dynamics of two levels with avoided crossing can be simply described with the help of so-called \textit{diabatic} basis formed by the instantaneous eigenfunction $\check{\Phi}^0_{kn}(t)$ of the time-dependent Hamiltonian (\ref{Hamiltonian_BdG_SOI}) at $\alpha=0$. Note here that for $\alpha=0$ the eigenstates do not depend of $h(t)$ at all and consist only of the Bogoliubov's amplitudes $u_0$ and $v_0$ (one can use (\ref{APP_A_eigenf}) and put $\alpha=0$ there).
The complete solution of the time-dependent hamiltonian can be written as follows
\begin{gather}\label{diab_basis}
\check{\Psi}_k(t)=\sum_nC^d_{kn}(t)\check{\Phi}^0_{kn}(t), 
\end{gather}
where $n=\{\uparrow+, \downarrow+, \uparrow-, \downarrow-\}$. 
In order to avoid confusion with adiabatic basis in (\ref{wf}) the superscript "d" is used to denote the diabatic basis.
The time-dependent coefficients obey the following equation derived from (\ref{Hamiltonian_BdG_SOI}):
\begin{gather}\label{eq1}
i \frac{\partial}{\partial t}C^d_m=
\sum_n C^d_n  \check{\Phi}^{0\dagger}_{km} \Big[\check{\mathcal{H}}(t)-i \frac{\partial}{\partial t} \Big] \check{\Phi}^0_{kn}.
\end{gather}
Note that here $\check{\mathcal{H}}(t)\check{\Phi}^0_{kn} \neq E_n(t) \check{\Phi}^0_{kn} $. 
By keeping in mind that $\hat{\Phi}^0_{kn}(t)$ depends on time only through $\Delta(t)$, one can rewrite (\ref{eq1}) as follows
\begin{widetext}
\begin{gather}\label{LZ_full}
i \frac{\partial}{\partial t}
\begin{pmatrix}
    C^d_{\uparrow +} \\
    C^d_{\downarrow +} \\
    C^d_{\uparrow -} \\
    C^d_{\downarrow -} 
\end{pmatrix}
=
\begin{pmatrix}
    {E_0-h(t)} &
    -\frac{\xi_k}{E_0}i\alpha k e^{-i\theta_k} &
    i\frac{\xi_k }{2E_0^2}\frac{\partial \Delta}{\partial t} &
    {\frac{\Delta}{E_0}i\alpha k e^{-i\theta_k}}  \\
    \frac{\xi_k}{E_0}i\alpha k e^{i\theta_k} & 
    E_0+h(t) & 
    -\frac{\Delta}{E_0}i\alpha k e^{i\theta_k} & 
    i\frac{\xi_k }{2E_0^2}\frac{\partial \Delta}{\partial t} \\
    -i\frac{\xi_k }{2E_0^2}\frac{\partial \Delta}{\partial t} & 
    \frac{\Delta}{E_0}i\alpha k e^{-i\theta_k}  & 
     -E_0-h(t) &
     \frac{\xi_k}{E_0}i\alpha k e^{-i\theta_k} \\
     {-\frac{\Delta}{E_0}i\alpha k e^{i\theta_k}} & 
     -i\frac{\xi_k }{2E_0^2}\frac{\partial \Delta}{\partial t} & 
     -\frac{\xi_k}{E_0}i\alpha k e^{i\theta_k} & 
     {-E_0+h(t)}
\end{pmatrix}
\begin{pmatrix}
    C^d_{\uparrow +} \\
    C^d_{\downarrow +} \\
    C^d_{\uparrow -} \\
    C^d_{\downarrow -} 
\end{pmatrix},
\end{gather}
where $E_0=\sqrt{\xi_k^2+\Delta^2}$.
One can remove the phase $\theta_k=\arg\big(k_x+ik_y\big)$ from (\ref{LZ_full}) by the unitary operator
\begin{gather}
\hat{U}_\theta=
\begin{pmatrix}
e^{i(\frac{\pi}{4}-\frac{\theta_k}{2})\hat{\sigma}_z} & 0    \\
0 & e^{i(\frac{\pi}{4}-\frac{\theta_k}{2})\hat{\sigma}_z}
\end{pmatrix},
\end{gather}
so that in the new basis we have
\begin{gather}\label{C_tilde_eq}
i \frac{\partial}{\partial t}
\begin{pmatrix}
    \tilde{C}^d_{\uparrow +} \\
    \tilde{C}^d_{\downarrow +} \\
    \tilde{C}^d_{\uparrow -} \\
    \tilde{C}^d_{\downarrow -} 
\end{pmatrix}
=
\begin{pmatrix}
    E_0-h(t) &
    -\frac{\xi}{E_0}\alpha k  &
    i\frac{\xi }{E_0^2}\frac{\partial \Delta}{\partial t} &
    \frac{\Delta}{E_0}\alpha k   \\
    -\frac{\xi}{E_0}\alpha k  & 
    E_0+h(t) & 
    \frac{\Delta}{E_0}\alpha k  & 
    i\frac{\xi }{E_0^2}\frac{\partial \Delta}{\partial t} \\
    -i\frac{\xi }{E_0^2}\frac{\partial \Delta}{\partial t} & 
    \frac{\Delta}{E_0}\alpha k   & 
     -E_0-h(t) &
     \frac{\xi}{E_0}\alpha k  \\
     \frac{\Delta}{E_0}\alpha k  & 
     -i\frac{\xi }{E_0^2}\frac{\partial \Delta}{\partial t} & 
     \frac{\xi}{E_0}\alpha k  & 
     -E_0+h(t)
\end{pmatrix}
\begin{pmatrix}
    \tilde{C}^d_{\uparrow +} \\
    \tilde{C}^d_{\downarrow +} \\
    \tilde{C}^d_{\uparrow -} \\
    \tilde{C}^d_{\downarrow -} 
\end{pmatrix}.
\end{gather}
\end{widetext}

We assume that the time evolution of the gap function $\Delta(t)$ is adiabatic on the timescale of the problem (\ref{C_tilde_eq}). Therefore one can assume $\Delta$ to be constant during the transition with the typical time $\sim\tau_\text{LZ}$.
Since the most emphasized dynamics occurs between two crossing branches, it is convenient to consider the interaction of only the corresponding terms $C_{k\uparrow+}$ and $C_{k\downarrow-}$ [Fig. \ref{fig3}]. 
Hence, one can extract an effective two-level problem for the crossing levels:
\begin{gather}\label{LZ_2l}
    i \frac{\partial}{\partial t}
\begin{pmatrix}
\tilde{C}^d_{k\uparrow+}       \\
\tilde{C}^d_{k\downarrow-}    
\end{pmatrix}=
\begin{pmatrix}
E_0-\gamma t & \frac{\Delta}{E_0}\alpha k    \\
\frac{\Delta}{E_0}\alpha k  & -E_0+\gamma t
\end{pmatrix}
\begin{pmatrix}
\tilde{C}^d_{k\uparrow+}       \\
\tilde{C}^d_{k\downarrow-}   
\end{pmatrix}.
\end{gather}

This system can be viewed as the LZSM problem, which allows an exact solution \cite{ivakhnenko_nonadiabatic_2023}.
However, as discussed in \ref{sec_4_B}, one can neglect the transient dynamics of the $C_k(t)$ coefficients in the gap equation (\ref{I_SCE}) and use the transition matrix approach instead. Thus, we need to obtain the relation between the long-time asymptotes of the functions $\tilde{C}^d_{k\uparrow\downarrow+}(t)$ before ($t_0-$) and after ($t_0+$) transition at the point $t_0(\xi_k)=\sqrt{\xi_k^2+\Delta^2}/\gamma$. Here we use short notations $(t_0\mp)\approx t_0\mp\tau_\text{LZ}/2$. 
The asymptotic solution of the problem (\ref{LZ_2l}) is well-known \cite{ivakhnenko_nonadiabatic_2023} and reads 
\begin{gather}
\begin{pmatrix}
\tilde{C}^d_{k\uparrow+}(t_0+)       \\
\tilde{C}^d_{k\downarrow-}(t_0+)    
\end{pmatrix}= \\ \notag
\begin{pmatrix}
\sqrt{p_k} & -\text{sgn}(\alpha) \sqrt{1-p_k}e^{i\chi_k} \\
\text{sgn}(\alpha)\sqrt{1-p_k}e^{-i\chi_k} & \sqrt{p_k}      
\end{pmatrix}
\\ \notag\times
\begin{pmatrix}
\tilde{C}^d_{k\uparrow+}(t_0-)       \\
\tilde{C}^d_{k\downarrow-}(t_0-)     
\end{pmatrix},
\end{gather}
where the coefficient $$p_k=\exp\Big[-\delta_{\text{LZ}}\frac{\Delta^2}{\xi_k^2+\Delta^2}\Big]$$ with $\delta_\text{LZ}=\pi \alpha^2k^2/\gamma\approx\pi \alpha^2k_F^2/\gamma$ defines the probability of tunneling. 
Here $\chi_k = \pi/4+ \text{arg} \Gamma ( 1 + i \ln p_k/2\pi ) - \ln p_k( \ln (-\ln p_k/2\pi) - 1 )/2\pi $ is the Stokes phase with the Gamma function $\Gamma$. 

For small energies $\xi_k \lesssim \Delta$ two different tunneling regimes are possible: 
\begin{gather} \notag
(\text{weak}) \quad \gamma \gtrsim \alpha^2 k_F^2 \quad \rightarrow \quad \delta_\text{LZ}\approx 0 \quad \rightarrow \quad p_k\approx 1 ,  \\ \notag
(\text{strong}) \quad \gamma \ll \alpha^2 k_F^2 \quad \rightarrow \quad \delta_\text{LZ}\gg 1  \quad \rightarrow \quad p_k\approx 0.
\end{gather}
When $\xi_k \gg \Delta$, tunneling is suppressed ($p_k\rightarrow 1$) as the quasiparticle spectrum resembles that of a normal metal with no splitting between crossing spectral branches.

The typical transient time $\tau_\text{LZ}$ for the LZSM tunneling can be estimated as follows \cite{ivakhnenko_nonadiabatic_2023}
$$\tau_{\text{LZ}}\sim \sqrt{\frac{\hbar}{\gamma}}\text{max}\Bigg\{1, \frac{\alpha k_F}{\sqrt{2\gamma}}\frac{\Delta}{\sqrt{\xi_k^2+\Delta^2}}\Bigg\}.$$
If the intersection of the branches of the QP spectrum occurs at some $\xi_k$, then it is possible to determine the interval $\Delta\xi_k$ in which all QP states experience transient dynamics. The size of $\Delta\xi_k$ depends on transient time, however, it can be shown, that the upper limit for this interval is $\Delta\xi_k \sim \alpha k_F \ll \Delta$.
The smalness of $\Delta\xi_k$ and the fact that the gap function $\Delta(t)$ is determined by all QP states in $(-\omega_D, \omega_D)$ confirm the validity of the approximations made in Section \ref{sec_4_B}.

Combining all the results we write the asymptotic transition matrix $\hat{S}^d_\text{LZ}$ in diabatic basis as
\begin{widetext}
\begin{gather}\label{slz}
\begin{pmatrix}
C^d_{k\uparrow+}(t_0+)       \\
C^d_{k\downarrow+}(t_0+)       \\
C^d_{k\uparrow-}(t_0+)       \\
C^d_{k\downarrow-}(t_0+)    
\end{pmatrix}=
\begin{pmatrix}
    \sqrt{p_k} & 0 & 0 &  \sqrt{1-p_k}e^{i\chi_k-i\theta_k-i\frac{\pi}{2}\text{sgn}(\alpha)} \\
0 & 1 & 0 & 0 \\
0 & 0 & 1 & 0 \\
-\sqrt{1-p_k}e^{-i\chi_k+i\theta_k+i\frac{\pi}{2}\text{sgn}(\alpha)} & 0 & 0 & \sqrt{p_k}     
\end{pmatrix}
\begin{pmatrix}
C^d_{k\uparrow+}(t_0-)       \\
C^d_{k\downarrow+}(t_0-)       \\
C^d_{k\uparrow-}(t_0-)       \\
C^d_{k\downarrow-}(t_0-)     
\end{pmatrix}.
\end{gather}
\end{widetext}

The LZSM transition matrix in the adiabatic basis (\ref{wf}) has the form $\hat{S}_\text{LZ}=\hat{R}^{-1}(t_0+)\hat{S}^d_\text{LZ}\hat{R}(t_0-)$, where we use the relationship between the two basis (\ref{diab_basis}) and (\ref{wf}) written in general form as a time-dependent matrix $\hat {R}(t)$. 
Using the perturbation theory with respect to the small parameter $\alpha k_F/\Delta$ and considering points $t_0\pm$ far from the nonadiabatic region, one can show that the matrix $\hat{R}(t_0\pm)$ can be approximated with an identity matrix. The corrections proportional to $\alpha k_F/\Delta$ in all elements of the matrix $\hat{R}(t_0\pm)$ as well as $\hat{S}_\text{LZ}$ can be neglected, since in all equations of Section \ref{sec_4} we consider the minimum possible order of the perturbation theory with respect to the parameter $\alpha k_F/\Delta$. With these approximations the matrices $\hat{S}_\text{LZ}$ and $\hat{S}^d_\text{LZ}$ actually coincide and the LZSM transition matrix in the adiabatic basis can be taken taken from (\ref{slz}). Thus, we get the Eq (\ref{S_lz}).

\begin{widetext}
\section{Calculation of spin-split DOS}\label{APP_E}

The DOS for one spin projection can be written as follows
\begin{gather}
N_\uparrow(E,t)\approx \sum_{k}\sum_{n=\uparrow+,\uparrow-} |u_0|^2\delta\Big(E-E_{kn}[h(t)]\Big) 
+|v_0|^2\delta\Big(E+E_{kn}[h(t)]\Big).
\end{gather}
Here we use static QP amplitudes $u_0$ and $v_0$ to distinguish the particle/hole contributions and $E_{k\uparrow\pm}$ are defined in Eq. (\ref{spectrum}). Note that the function $N_{\uparrow}(E,t)$ depends on time only through the Zeeman field $h(t)$. 
The straightforward calculations for $h(t)<\Delta_0$ yield
\begin{gather}\label{DOS_1}
\frac{N_\uparrow(E,t)}{N(0)}\approx 
\frac{|E|}{\xi_0}\Bigg| 1-\text{sgn}(E)\frac{\alpha^2 k_F^2+h^2(t)}{\sqrt{\xi_0^2\alpha^2k^2_F+h^2(t)(\xi_0^2+\Delta_h^2)}} \Bigg|^{-1},
\end{gather}
where
\begin{gather}
\xi_{0}(E,t)\approx  
\sqrt{ E^2+h^2(t)-\Delta_h^2+ \alpha^2k_F^2 +\text{sgn}(E) 2\sqrt{E^2(h^2(t)+ \alpha^2k_F^2)-\Delta_h^2 \alpha^2k_F^2} }
\end{gather} 
and we have assumed $\alpha k\approx\alpha k_F$ due to the vicinity to the Fermi energy. The time-dependent gap function $\Delta_h[h(t)]$ is defined in (\ref{D_h(t)}). Two standard coherence peaks at the energies $E = -\sqrt{(\Delta_h+h(t))^2+\alpha^2k_F^2}$ and $E = \sqrt{(\Delta_h-h(t))^2+\alpha^2k_F^2}$ appear [Fig. \ref{fig6}(a)].

For the case of large Zeeman fields $h(t)>\Delta_0$ one obtains
\begin{gather}\label{DOS_2}
\frac{N_\uparrow(E,t)}{N(0)}\approx 
\begin{cases}
\frac{|E|}{\xi_0}\Bigg| 1-\text{sgn}(E)\frac{\alpha^2 k_F^2+h^2(t)}{\sqrt{\xi_1^2\alpha^2k^2_F+h^2(t)(\xi_1^2+\Delta_h^2)}} \Bigg|^{-1}, \quad E>\Delta_m \quad \text{and} \quad  E<-\sqrt{(\Delta_h+h(t))^2+\alpha^2k_F^2}
\\ 
\frac{|E|}{\xi_0}\Bigg| 1-\frac{\alpha^2 k_F^2+h^2(t)}{\sqrt{\xi_2^2\alpha^2k^2_F+h^2(t)(\xi_2^2+\Delta_h^2)}} \Bigg|^{-1}, \quad -\sqrt{(\Delta_h-h(t))^2+\alpha^2k_F^2}< E<-\Delta_m
\end{cases}.
\end{gather}
where 
$$ \Delta_m \approx \frac{\Delta_h \alpha k_F}{\sqrt{h^2(t)+\alpha^2k_F^2}}. $$
The splitting of the energy spectrum in the vicinity of $E=0$ leads to the appearance of the two additional coherence peaks and corresponging minigap at the energies $E=\pm\Delta_m$, which are shown in Fig. \ref{fig6}(b). 
\begin{figure}[] 
\centering
\begin{minipage}[h]{0.6\linewidth}
\includegraphics[width=1.0\textwidth]{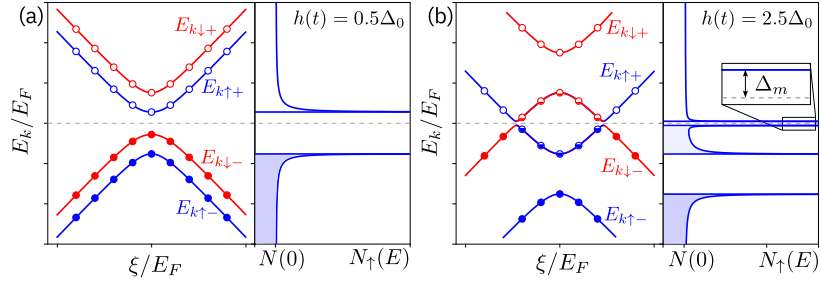} 
\end{minipage}
\caption{ {\small Spectrum (\ref{spectrum}) and density of states (\ref{DOS_1}- \ref{DOS_2}) for the QPs with $\uparrow$ spin for two different values of $h(t)$. The value $\Delta_m$ represents a minigap. Colored areas in DOS indicate the filling  of the states in the corresponding energy intervals according to Eq. (\ref{distrib}).
The parameters are chosen as in Fig. (\ref{fig3}): $\Delta_0/E_F=0.01$, $\alpha/E_F=0.0025$. }}
\label{fig6}
\end{figure}

\end{widetext}


\begin{thebibliography}{48}%
\makeatletter
\providecommand \@ifxundefined [1]{%
 \@ifx{#1\undefined}
}%
\providecommand \@ifnum [1]{%
 \ifnum #1\expandafter \@firstoftwo
 \else \expandafter \@secondoftwo
 \fi
}%
\providecommand \@ifx [1]{%
 \ifx #1\expandafter \@firstoftwo
 \else \expandafter \@secondoftwo
 \fi
}%
\providecommand \natexlab [1]{#1}%
\providecommand \enquote  [1]{``#1''}%
\providecommand \bibnamefont  [1]{#1}%
\providecommand \bibfnamefont [1]{#1}%
\providecommand \citenamefont [1]{#1}%
\providecommand \href@noop [0]{\@secondoftwo}%
\providecommand \href [0]{\begingroup \@sanitize@url \@href}%
\providecommand \@href[1]{\@@startlink{#1}\@@href}%
\providecommand \@@href[1]{\endgroup#1\@@endlink}%
\providecommand \@sanitize@url [0]{\catcode `\\12\catcode `\$12\catcode `\&12\catcode `\#12\catcode `\^12\catcode `\_12\catcode `\%12\relax}%
\providecommand \@@startlink[1]{}%
\providecommand \@@endlink[0]{}%
\providecommand \url  [0]{\begingroup\@sanitize@url \@url }%
\providecommand \@url [1]{\endgroup\@href {#1}{\urlprefix }}%
\providecommand \urlprefix  [0]{URL }%
\providecommand \Eprint [0]{\href }%
\providecommand \doibase [0]{http://dx.doi.org/}%
\providecommand \selectlanguage [0]{\@gobble}%
\providecommand \bibinfo  [0]{\@secondoftwo}%
\providecommand \bibfield  [0]{\@secondoftwo}%
\providecommand \translation [1]{[#1]}%
\providecommand \BibitemOpen [0]{}%
\providecommand \bibitemStop [0]{}%
\providecommand \bibitemNoStop [0]{.\EOS\space}%
\providecommand \EOS [0]{\spacefactor3000\relax}%
\providecommand \BibitemShut  [1]{\csname bibitem#1\endcsname}%
\let\auto@bib@innerbib\@empty
\bibitem [{\citenamefont {Langenberg}\ \emph {et~al.}(1986)\citenamefont {Langenberg}, \citenamefont {Larkin},\ and\ \citenamefont {Larkin}}]{langenberg1986nonequilibrium}%
  \BibitemOpen
  \bibfield  {author} {\bibinfo {author} {\bibfnamefont {D.}~\bibnamefont {Langenberg}}, \bibinfo {author} {\bibfnamefont {A.}~\bibnamefont {Larkin}}, \ and\ \bibinfo {author} {\bibfnamefont {A.}~\bibnamefont {Larkin}},\ }\href {https://books.google.fr/books?id=alh6AAAAIAAJ} {\emph {\bibinfo {title} {Nonequilibrium Superconductivity}}},\ Modern problems in condensed matter sciences\ (\bibinfo  {publisher} {North-Holland},\ \bibinfo {year} {1986})\BibitemShut {NoStop}%
\bibitem [{\citenamefont {Kopnin}(2001)}]{kopnin2001theory}%
  \BibitemOpen
  \bibfield  {author} {\bibinfo {author} {\bibfnamefont {N.}~\bibnamefont {Kopnin}},\ }\href {https://books.google.fr/books?id=VuA-DgAAQBAJ} {\emph {\bibinfo {title} {Theory of Nonequilibrium Superconductivity}}},\ International Series of Monographs on Physics\ (\bibinfo  {publisher} {Clarendon Press},\ \bibinfo {year} {2001})\BibitemShut {NoStop}%
\bibitem [{\citenamefont {{Volkov}}\ and\ \citenamefont {{Kogan}}(1974)}]{1974JETP38.1018V}%
  \BibitemOpen
  \bibfield  {author} {\bibinfo {author} {\bibfnamefont {A.~F.}\ \bibnamefont {{Volkov}}}\ and\ \bibinfo {author} {\bibfnamefont {S.~M.}\ \bibnamefont {{Kogan}}},\ }\bibfield  {title} {\bibinfo {title} {{Collisionless relaxation of the energy gap in superconductors}},\ }\href@noop {} {\bibfield  {journal} {\bibinfo  {journal} {Soviet Journal of Experimental and Theoretical Physics}\ }\textbf {\bibinfo {volume} {38}},\ \bibinfo {pages} {1018} (\bibinfo {year} {1974})}\BibitemShut {NoStop}%
\bibitem [{\citenamefont {Kulik}\ \emph {et~al.}(1981)\citenamefont {Kulik}, \citenamefont {Entin-Wohlman},\ and\ \citenamefont {Orbach}}]{kulik_pair_1981}%
  \BibitemOpen
  \bibfield  {author} {\bibinfo {author} {\bibfnamefont {I.~O.}\ \bibnamefont {Kulik}}, \bibinfo {author} {\bibfnamefont {O.}~\bibnamefont {Entin-Wohlman}}, \ and\ \bibinfo {author} {\bibfnamefont {R.}~\bibnamefont {Orbach}},\ }\bibfield  {title} {\bibinfo {title} {Pair susceptibility and mode propagation in superconductors: A microscopic approach},\ }\href {\doibase 10.1007/BF00115617} {\bibfield  {journal} {\bibinfo  {journal} {Journal of Low Temperature Physics}\ }\textbf {\bibinfo {volume} {43}},\ \bibinfo {pages} {591} (\bibinfo {year} {1981})}\BibitemShut {NoStop}%
\bibitem [{\citenamefont {Higgs}(1964)}]{PhysRevLett.13.508}%
  \BibitemOpen
  \bibfield  {author} {\bibinfo {author} {\bibfnamefont {P.~W.}\ \bibnamefont {Higgs}},\ }\bibfield  {title} {\bibinfo {title} {Broken symmetries and the masses of gauge bosons},\ }\href {\doibase 10.1103/PhysRevLett.13.508} {\bibfield  {journal} {\bibinfo  {journal} {Phys. Rev. Lett.}\ }\textbf {\bibinfo {volume} {13}},\ \bibinfo {pages} {508} (\bibinfo {year} {1964})}\BibitemShut {NoStop}%
\bibitem [{\citenamefont {Pekker}\ and\ \citenamefont {Varma}(2015)}]{pekker_amplitudehiggs_2015}%
  \BibitemOpen
  \bibfield  {author} {\bibinfo {author} {\bibfnamefont {D.}~\bibnamefont {Pekker}}\ and\ \bibinfo {author} {\bibfnamefont {C.}~\bibnamefont {Varma}},\ }\bibfield  {title} {\bibinfo {title} {Amplitude/{H}iggs modes in condensed matter physics},\ }\href {\doibase 10.1146/annurev-conmatphys-031214-014350} {\bibfield  {journal} {\bibinfo  {journal} {Annual Review of Condensed Matter Physics}\ }\textbf {\bibinfo {volume} {6}},\ \bibinfo {pages} {269} (\bibinfo {year} {2015})}\BibitemShut {NoStop}%
\bibitem [{\citenamefont {Moor}\ \emph {et~al.}(2017)\citenamefont {Moor}, \citenamefont {Volkov},\ and\ \citenamefont {Efetov}}]{PhysRevLett.118.047001}%
  \BibitemOpen
  \bibfield  {author} {\bibinfo {author} {\bibfnamefont {A.}~\bibnamefont {Moor}}, \bibinfo {author} {\bibfnamefont {A.~F.}\ \bibnamefont {Volkov}}, \ and\ \bibinfo {author} {\bibfnamefont {K.~B.}\ \bibnamefont {Efetov}},\ }\bibfield  {title} {\bibinfo {title} {Amplitude {H}iggs mode and admittance in superconductors with a moving condensate},\ }\href {\doibase 10.1103/PhysRevLett.118.047001} {\bibfield  {journal} {\bibinfo  {journal} {Phys. Rev. Lett.}\ }\textbf {\bibinfo {volume} {118}},\ \bibinfo {pages} {047001} (\bibinfo {year} {2017})}\BibitemShut {NoStop}%
\bibitem [{\citenamefont {Nakamura}\ \emph {et~al.}(2019)\citenamefont {Nakamura}, \citenamefont {Iida}, \citenamefont {Murotani}, \citenamefont {Matsunaga}, \citenamefont {Terai},\ and\ \citenamefont {Shimano}}]{PhysRevLett.122.257001}%
  \BibitemOpen
  \bibfield  {author} {\bibinfo {author} {\bibfnamefont {S.}~\bibnamefont {Nakamura}}, \bibinfo {author} {\bibfnamefont {Y.}~\bibnamefont {Iida}}, \bibinfo {author} {\bibfnamefont {Y.}~\bibnamefont {Murotani}}, \bibinfo {author} {\bibfnamefont {R.}~\bibnamefont {Matsunaga}}, \bibinfo {author} {\bibfnamefont {H.}~\bibnamefont {Terai}}, \ and\ \bibinfo {author} {\bibfnamefont {R.}~\bibnamefont {Shimano}},\ }\bibfield  {title} {\bibinfo {title} {Infrared activation of the {H}iggs mode by supercurrent injection in superconducting {NbN}},\ }\href {\doibase 10.1103/PhysRevLett.122.257001} {\bibfield  {journal} {\bibinfo  {journal} {Phys. Rev. Lett.}\ }\textbf {\bibinfo {volume} {122}},\ \bibinfo {pages} {257001} (\bibinfo {year} {2019})}\BibitemShut {NoStop}%
\bibitem [{\citenamefont {Bellitti}\ \emph {et~al.}(2022)\citenamefont {Bellitti}, \citenamefont {Laumann},\ and\ \citenamefont {Spivak}}]{PhysRevB.105.104513}%
  \BibitemOpen
  \bibfield  {author} {\bibinfo {author} {\bibfnamefont {M.}~\bibnamefont {Bellitti}}, \bibinfo {author} {\bibfnamefont {C.~R.}\ \bibnamefont {Laumann}}, \ and\ \bibinfo {author} {\bibfnamefont {B.~Z.}\ \bibnamefont {Spivak}},\ }\bibfield  {title} {\bibinfo {title} {Incoherent excitation of coherent {H}iggs oscillations in superconductors},\ }\href {\doibase 10.1103/PhysRevB.105.104513} {\bibfield  {journal} {\bibinfo  {journal} {Phys. Rev. B}\ }\textbf {\bibinfo {volume} {105}},\ \bibinfo {pages} {104513} (\bibinfo {year} {2022})}\BibitemShut {NoStop}%
\bibitem [{\citenamefont {Papenkort}\ \emph {et~al.}(2007)\citenamefont {Papenkort}, \citenamefont {Axt},\ and\ \citenamefont {Kuhn}}]{PhysRevB.76.224522}%
  \BibitemOpen
  \bibfield  {author} {\bibinfo {author} {\bibfnamefont {T.}~\bibnamefont {Papenkort}}, \bibinfo {author} {\bibfnamefont {V.~M.}\ \bibnamefont {Axt}}, \ and\ \bibinfo {author} {\bibfnamefont {T.}~\bibnamefont {Kuhn}},\ }\bibfield  {title} {\bibinfo {title} {Coherent dynamics and pump-probe spectra of {BCS} superconductors},\ }\href {\doibase 10.1103/PhysRevB.76.224522} {\bibfield  {journal} {\bibinfo  {journal} {Phys. Rev. B}\ }\textbf {\bibinfo {volume} {76}},\ \bibinfo {pages} {224522} (\bibinfo {year} {2007})}\BibitemShut {NoStop}%
\bibitem [{\citenamefont {Matsunaga}\ \emph {et~al.}(2014)\citenamefont {Matsunaga}, \citenamefont {Tsuji}, \citenamefont {Fujita}, \citenamefont {Sugioka}, \citenamefont {Makise}, \citenamefont {Uzawa}, \citenamefont {Terai}, \citenamefont {Wang}, \citenamefont {Aoki},\ and\ \citenamefont {Shimano}}]{doi:10.1126/science.1254697}%
  \BibitemOpen
  \bibfield  {author} {\bibinfo {author} {\bibfnamefont {R.}~\bibnamefont {Matsunaga}}, \bibinfo {author} {\bibfnamefont {N.}~\bibnamefont {Tsuji}}, \bibinfo {author} {\bibfnamefont {H.}~\bibnamefont {Fujita}}, \bibinfo {author} {\bibfnamefont {A.}~\bibnamefont {Sugioka}}, \bibinfo {author} {\bibfnamefont {K.}~\bibnamefont {Makise}}, \bibinfo {author} {\bibfnamefont {Y.}~\bibnamefont {Uzawa}}, \bibinfo {author} {\bibfnamefont {H.}~\bibnamefont {Terai}}, \bibinfo {author} {\bibfnamefont {Z.}~\bibnamefont {Wang}}, \bibinfo {author} {\bibfnamefont {H.}~\bibnamefont {Aoki}}, \ and\ \bibinfo {author} {\bibfnamefont {R.}~\bibnamefont {Shimano}},\ }\bibfield  {title} {\bibinfo {title} {Light-induced collective pseudospin precession resonating with {H}iggs mode in a superconductor},\ }\href {\doibase 10.1126/science.1254697} {\bibfield  {journal} {\bibinfo  {journal} {Science}\ }\textbf {\bibinfo {volume} {345}},\ \bibinfo {pages} {1145} (\bibinfo {year} {2014})}\BibitemShut {NoStop}%
\bibitem [{\citenamefont {Matsunaga}\ \emph {et~al.}(2013)\citenamefont {Matsunaga}, \citenamefont {Hamada}, \citenamefont {Makise}, \citenamefont {Uzawa}, \citenamefont {Terai}, \citenamefont {Wang},\ and\ \citenamefont {Shimano}}]{PhysRevLett.111.057002}%
  \BibitemOpen
  \bibfield  {author} {\bibinfo {author} {\bibfnamefont {R.}~\bibnamefont {Matsunaga}}, \bibinfo {author} {\bibfnamefont {Y.~I.}\ \bibnamefont {Hamada}}, \bibinfo {author} {\bibfnamefont {K.}~\bibnamefont {Makise}}, \bibinfo {author} {\bibfnamefont {Y.}~\bibnamefont {Uzawa}}, \bibinfo {author} {\bibfnamefont {H.}~\bibnamefont {Terai}}, \bibinfo {author} {\bibfnamefont {Z.}~\bibnamefont {Wang}}, \ and\ \bibinfo {author} {\bibfnamefont {R.}~\bibnamefont {Shimano}},\ }\bibfield  {title} {\bibinfo {title} {Higgs amplitude mode in the bcs superconductors $\text{Nb}_{1-x}\text{Ti}_x\text{N}$ induced by terahertz pulse excitation},\ }\href {\doibase 10.1103/PhysRevLett.111.057002} {\bibfield  {journal} {\bibinfo  {journal} {Phys. Rev. Lett.}\ }\textbf {\bibinfo {volume} {111}},\ \bibinfo {pages} {057002} (\bibinfo {year} {2013})}\BibitemShut {NoStop}%
\bibitem [{\citenamefont {Kemper}\ \emph {et~al.}(2015)\citenamefont {Kemper}, \citenamefont {Sentef}, \citenamefont {Moritz}, \citenamefont {Freericks},\ and\ \citenamefont {Devereaux}}]{PhysRevB.92.224517}%
  \BibitemOpen
  \bibfield  {author} {\bibinfo {author} {\bibfnamefont {A.~F.}\ \bibnamefont {Kemper}}, \bibinfo {author} {\bibfnamefont {M.~A.}\ \bibnamefont {Sentef}}, \bibinfo {author} {\bibfnamefont {B.}~\bibnamefont {Moritz}}, \bibinfo {author} {\bibfnamefont {J.~K.}\ \bibnamefont {Freericks}}, \ and\ \bibinfo {author} {\bibfnamefont {T.~P.}\ \bibnamefont {Devereaux}},\ }\bibfield  {title} {\bibinfo {title} {Direct observation of {H}iggs mode oscillations in the pump-probe photoemission spectra of electron-phonon mediated superconductors},\ }\href {\doibase 10.1103/PhysRevB.92.224517} {\bibfield  {journal} {\bibinfo  {journal} {Phys. Rev. B}\ }\textbf {\bibinfo {volume} {92}},\ \bibinfo {pages} {224517} (\bibinfo {year} {2015})}\BibitemShut {NoStop}%
\bibitem [{\citenamefont {Shimano}\ and\ \citenamefont {Tsuji}(2020)}]{doi:10.1146/annurev-conmatphys-031119-050813}%
  \BibitemOpen
  \bibfield  {author} {\bibinfo {author} {\bibfnamefont {R.}~\bibnamefont {Shimano}}\ and\ \bibinfo {author} {\bibfnamefont {N.}~\bibnamefont {Tsuji}},\ }\bibfield  {title} {\bibinfo {title} {Higgs mode in superconductors},\ }\href {\doibase 10.1146/annurev-conmatphys-031119-050813} {\bibfield  {journal} {\bibinfo  {journal} {Annual Review of Condensed Matter Physics}\ }\textbf {\bibinfo {volume} {11}},\ \bibinfo {pages} {103} (\bibinfo {year} {2020})}\BibitemShut {NoStop}%
\bibitem [{\citenamefont {Buzdin}(2005)}]{buzdin_proximity_2005}%
  \BibitemOpen
  \bibfield  {author} {\bibinfo {author} {\bibfnamefont {A.~I.}\ \bibnamefont {Buzdin}},\ }\bibfield  {title} {\bibinfo {title} {Proximity effects in superconductor-ferromagnet heterostructures},\ }\href {\doibase 10.1103/RevModPhys.77.935} {\bibfield  {journal} {\bibinfo  {journal} {Reviews of Modern Physics}\ }\textbf {\bibinfo {volume} {77}},\ \bibinfo {pages} {935} (\bibinfo {year} {2005})}\BibitemShut {NoStop}%
\bibitem [{\citenamefont {Eschrig}(2015)}]{eschrig_spin-polarized_2015}%
  \BibitemOpen
  \bibfield  {author} {\bibinfo {author} {\bibfnamefont {M.}~\bibnamefont {Eschrig}},\ }\bibfield  {title} {\bibinfo {title} {Spin-polarized supercurrents for spintronics: a review of current progress},\ }\href {\doibase 10.1088/0034-4885/78/10/104501} {\bibfield  {journal} {\bibinfo  {journal} {Reports on Progress in Physics}\ }\textbf {\bibinfo {volume} {78}},\ \bibinfo {pages} {104501} (\bibinfo {year} {2015})}\BibitemShut {NoStop}%
\bibitem [{\citenamefont {Heikkilä}\ \emph {et~al.}(2019)\citenamefont {Heikkilä}, \citenamefont {Silaev}, \citenamefont {Virtanen},\ and\ \citenamefont {Bergeret}}]{heikkila_thermal_2019}%
  \BibitemOpen
  \bibfield  {author} {\bibinfo {author} {\bibfnamefont {T.~T.}\ \bibnamefont {Heikkilä}}, \bibinfo {author} {\bibfnamefont {M.}~\bibnamefont {Silaev}}, \bibinfo {author} {\bibfnamefont {P.}~\bibnamefont {Virtanen}}, \ and\ \bibinfo {author} {\bibfnamefont {F.~S.}\ \bibnamefont {Bergeret}},\ }\bibfield  {title} {\bibinfo {title} {Thermal, electric and spin transport in superconductor/ferromagnetic-insulator structures},\ }\href {\doibase 10.1016/j.progsurf.2019.100540} {\bibfield  {journal} {\bibinfo  {journal} {Progress in Surface Science}\ }\textbf {\bibinfo {volume} {94}},\ \bibinfo {pages} {100540} (\bibinfo {year} {2019})}\BibitemShut {NoStop}%
\bibitem [{\citenamefont {Houzet}(2008)}]{houzet_ferromagnetic_2008}%
  \BibitemOpen
  \bibfield  {author} {\bibinfo {author} {\bibfnamefont {M.}~\bibnamefont {Houzet}},\ }\bibfield  {title} {\bibinfo {title} {Ferromagnetic josephson junction with precessing magnetization},\ }\href {\doibase 10.1103/PhysRevLett.101.057009} {\bibfield  {journal} {\bibinfo  {journal} {Physical Review Letters}\ }\textbf {\bibinfo {volume} {101}},\ \bibinfo {pages} {057009} (\bibinfo {year} {2008})}\BibitemShut {NoStop}%
\bibitem [{\citenamefont {Barnes}\ \emph {et~al.}(2011)\citenamefont {Barnes}, \citenamefont {Aprili}, \citenamefont {Petković},\ and\ \citenamefont {Maekawa}}]{Barnes_2011}%
  \BibitemOpen
  \bibfield  {author} {\bibinfo {author} {\bibfnamefont {S.~E.}\ \bibnamefont {Barnes}}, \bibinfo {author} {\bibfnamefont {M.}~\bibnamefont {Aprili}}, \bibinfo {author} {\bibfnamefont {I.}~\bibnamefont {Petković}}, \ and\ \bibinfo {author} {\bibfnamefont {S.}~\bibnamefont {Maekawa}},\ }\bibfield  {title} {\bibinfo {title} {Ferromagnetic resonance with a magnetic josephson junction},\ }\href {\doibase 10.1088/0953-2048/24/2/024020} {\bibfield  {journal} {\bibinfo  {journal} {Superconductor Science and Technology}\ }\textbf {\bibinfo {volume} {24}},\ \bibinfo {pages} {024020} (\bibinfo {year} {2011})}\BibitemShut {NoStop}%
\bibitem [{\citenamefont {Petkovi\ifmmode~\acute{c}\else \'{c}\fi{}}\ \emph {et~al.}(2009)\citenamefont {Petkovi\ifmmode~\acute{c}\else \'{c}\fi{}}, \citenamefont {Aprili}, \citenamefont {Barnes}, \citenamefont {Beuneu},\ and\ \citenamefont {Maekawa}}]{PhysRevB.80.220502}%
  \BibitemOpen
  \bibfield  {author} {\bibinfo {author} {\bibfnamefont {I.}~\bibnamefont {Petkovi\ifmmode~\acute{c}\else \'{c}\fi{}}}, \bibinfo {author} {\bibfnamefont {M.}~\bibnamefont {Aprili}}, \bibinfo {author} {\bibfnamefont {S.~E.}\ \bibnamefont {Barnes}}, \bibinfo {author} {\bibfnamefont {F.}~\bibnamefont {Beuneu}}, \ and\ \bibinfo {author} {\bibfnamefont {S.}~\bibnamefont {Maekawa}},\ }\bibfield  {title} {\bibinfo {title} {Direct dynamical coupling of spin modes and singlet josephson supercurrent in ferromagnetic josephson junctions},\ }\href {\doibase 10.1103/PhysRevB.80.220502} {\bibfield  {journal} {\bibinfo  {journal} {Phys. Rev. B}\ }\textbf {\bibinfo {volume} {80}},\ \bibinfo {pages} {220502} (\bibinfo {year} {2009})}\BibitemShut {NoStop}%
\bibitem [{\citenamefont {Takahashi}\ \emph {et~al.}(2007)\citenamefont {Takahashi}, \citenamefont {Hikino}, \citenamefont {Mori}, \citenamefont {Martinek},\ and\ \citenamefont {Maekawa}}]{PhysRevLett.99.057003}%
  \BibitemOpen
  \bibfield  {author} {\bibinfo {author} {\bibfnamefont {S.}~\bibnamefont {Takahashi}}, \bibinfo {author} {\bibfnamefont {S.}~\bibnamefont {Hikino}}, \bibinfo {author} {\bibfnamefont {M.}~\bibnamefont {Mori}}, \bibinfo {author} {\bibfnamefont {J.}~\bibnamefont {Martinek}}, \ and\ \bibinfo {author} {\bibfnamefont {S.}~\bibnamefont {Maekawa}},\ }\bibfield  {title} {\bibinfo {title} {Supercurrent pumping in josephson junctions with a half-metallic ferromagnet},\ }\href {\doibase 10.1103/PhysRevLett.99.057003} {\bibfield  {journal} {\bibinfo  {journal} {Phys. Rev. Lett.}\ }\textbf {\bibinfo {volume} {99}},\ \bibinfo {pages} {057003} (\bibinfo {year} {2007})}\BibitemShut {NoStop}%
\bibitem [{\citenamefont {Li}\ \emph {et~al.}(2018)\citenamefont {Li}, \citenamefont {Zhao}, \citenamefont {Zhang},\ and\ \citenamefont {Sun}}]{Li_2018}%
  \BibitemOpen
  \bibfield  {author} {\bibinfo {author} {\bibfnamefont {L.-L.}\ \bibnamefont {Li}}, \bibinfo {author} {\bibfnamefont {Y.-L.}\ \bibnamefont {Zhao}}, \bibinfo {author} {\bibfnamefont {X.-X.}\ \bibnamefont {Zhang}}, \ and\ \bibinfo {author} {\bibfnamefont {Y.}~\bibnamefont {Sun}},\ }\bibfield  {title} {\bibinfo {title} {Possible evidence for spin-transfer torque induced by spin-triplet supercurrents*},\ }\href {\doibase 10.1088/0256-307X/35/7/077401} {\bibfield  {journal} {\bibinfo  {journal} {Chinese Physics Letters}\ }\textbf {\bibinfo {volume} {35}},\ \bibinfo {pages} {077401} (\bibinfo {year} {2018})}\BibitemShut {NoStop}%
\bibitem [{\citenamefont {Golovchanskiy}\ \emph {et~al.}(2023)\citenamefont {Golovchanskiy}, \citenamefont {Abramov}, \citenamefont {Emelyanova}, \citenamefont {Shchetinin}, \citenamefont {Ryazanov}, \citenamefont {Golubov},\ and\ \citenamefont {Stolyarov}}]{PhysRevApplied.19.034025}%
  \BibitemOpen
  \bibfield  {author} {\bibinfo {author} {\bibfnamefont {I.}~\bibnamefont {Golovchanskiy}}, \bibinfo {author} {\bibfnamefont {N.}~\bibnamefont {Abramov}}, \bibinfo {author} {\bibfnamefont {O.}~\bibnamefont {Emelyanova}}, \bibinfo {author} {\bibfnamefont {I.}~\bibnamefont {Shchetinin}}, \bibinfo {author} {\bibfnamefont {V.}~\bibnamefont {Ryazanov}}, \bibinfo {author} {\bibfnamefont {A.}~\bibnamefont {Golubov}}, \ and\ \bibinfo {author} {\bibfnamefont {V.}~\bibnamefont {Stolyarov}},\ }\bibfield  {title} {\bibinfo {title} {Magnetization dynamics in proximity-coupled superconductor-ferromagnet-superconductor multilayers. ii. thickness dependence of the superconducting torque},\ }\href {\doibase 10.1103/PhysRevApplied.19.034025} {\bibfield  {journal} {\bibinfo  {journal} {Phys. Rev. Appl.}\ }\textbf {\bibinfo {volume} {19}},\ \bibinfo {pages} {034025} (\bibinfo {year} {2023})}\BibitemShut {NoStop}%
\bibitem [{\citenamefont {Silaev}(2022)}]{PhysRevApplied.18.L061004}%
  \BibitemOpen
  \bibfield  {author} {\bibinfo {author} {\bibfnamefont {M.}~\bibnamefont {Silaev}},\ }\bibfield  {title} {\bibinfo {title} {Anderson-$\text{H}$iggs mass of magnons in superconductor-ferromagnet-superconductor systems},\ }\href {\doibase 10.1103/PhysRevApplied.18.L061004} {\bibfield  {journal} {\bibinfo  {journal} {Phys. Rev. Appl.}\ }\textbf {\bibinfo {volume} {18}},\ \bibinfo {pages} {L061004} (\bibinfo {year} {2022})}\BibitemShut {NoStop}%
\bibitem [{\citenamefont {Vadimov}\ \emph {et~al.}(2019)\citenamefont {Vadimov}, \citenamefont {Khaymovich},\ and\ \citenamefont {Mel'nikov}}]{vadimov_higgs_2019}%
  \BibitemOpen
  \bibfield  {author} {\bibinfo {author} {\bibfnamefont {V.~L.}\ \bibnamefont {Vadimov}}, \bibinfo {author} {\bibfnamefont {I.~M.}\ \bibnamefont {Khaymovich}}, \ and\ \bibinfo {author} {\bibfnamefont {A.~S.}\ \bibnamefont {Mel'nikov}},\ }\bibfield  {title} {\bibinfo {title} {Higgs modes in proximized superconducting systems},\ }\href {\doibase 10.1103/PhysRevB.100.104515} {\bibfield  {journal} {\bibinfo  {journal} {Physical Review B}\ }\textbf {\bibinfo {volume} {100}},\ \bibinfo {pages} {104515} (\bibinfo {year} {2019})}\BibitemShut {NoStop}%
\bibitem [{\citenamefont {Tang}\ \emph {et~al.}(2020)\citenamefont {Tang}, \citenamefont {Belzig}, \citenamefont {Z\"ulicke},\ and\ \citenamefont {Bruder}}]{PhysRevResearch.2.022068}%
  \BibitemOpen
  \bibfield  {author} {\bibinfo {author} {\bibfnamefont {G.}~\bibnamefont {Tang}}, \bibinfo {author} {\bibfnamefont {W.}~\bibnamefont {Belzig}}, \bibinfo {author} {\bibfnamefont {U.}~\bibnamefont {Z\"ulicke}}, \ and\ \bibinfo {author} {\bibfnamefont {C.}~\bibnamefont {Bruder}},\ }\bibfield  {title} {\bibinfo {title} {Signatures of the {H}iggs mode in transport through a normal-metal--superconductor junction},\ }\href {\doibase 10.1103/PhysRevResearch.2.022068} {\bibfield  {journal} {\bibinfo  {journal} {Phys. Rev. Res.}\ }\textbf {\bibinfo {volume} {2}},\ \bibinfo {pages} {022068} (\bibinfo {year} {2020})}\BibitemShut {NoStop}%
\bibitem [{\citenamefont {Lu}\ \emph {et~al.}(2022{\natexlab{a}})\citenamefont {Lu}, \citenamefont {Ilić}, \citenamefont {Ojajärvi}, \citenamefont {Heikkilä},\ and\ \citenamefont {Bergeret}}]{https://doi.org/10.48550/arxiv.2212.11615}%
  \BibitemOpen
  \bibfield  {author} {\bibinfo {author} {\bibfnamefont {Y.}~\bibnamefont {Lu}}, \bibinfo {author} {\bibfnamefont {S.}~\bibnamefont {Ilić}}, \bibinfo {author} {\bibfnamefont {R.}~\bibnamefont {Ojajärvi}}, \bibinfo {author} {\bibfnamefont {T.~T.}\ \bibnamefont {Heikkilä}}, \ and\ \bibinfo {author} {\bibfnamefont {F.~S.}\ \bibnamefont {Bergeret}},\ }\bibfield  {title} {\bibinfo {title} {Reducing the frequency of the {H}iggs mode in a helical superconductor coupled to an $\text{LC}$-circuit},\ }\href {https://arxiv.org/abs/2212.11615} {\bibfield  {journal} {\bibinfo  {journal} {arXiv:2212.11615}\ } (\bibinfo {year} {2022}{\natexlab{a}})}\BibitemShut {NoStop}%
\bibitem [{\citenamefont {Silaev}\ \emph {et~al.}(2020)\citenamefont {Silaev}, \citenamefont {Ojajärvi},\ and\ \citenamefont {Heikkilä}}]{silaev_spin_2020}%
  \BibitemOpen
  \bibfield  {author} {\bibinfo {author} {\bibfnamefont {M.~A.}\ \bibnamefont {Silaev}}, \bibinfo {author} {\bibfnamefont {R.}~\bibnamefont {Ojajärvi}}, \ and\ \bibinfo {author} {\bibfnamefont {T.~T.}\ \bibnamefont {Heikkilä}},\ }\bibfield  {title} {\bibinfo {title} {Spin and charge currents driven by the higgs mode in high-field superconductors},\ }\href {\doibase 10.1103/PhysRevResearch.2.033416} {\bibfield  {journal} {\bibinfo  {journal} {Physical Review Research}\ }\textbf {\bibinfo {volume} {2}},\ \bibinfo {pages} {033416} (\bibinfo {year} {2020})}\BibitemShut {NoStop}%
\bibitem [{\citenamefont {Lu}\ \emph {et~al.}(2022{\natexlab{b}})\citenamefont {Lu}, \citenamefont {Ojajärvi}, \citenamefont {Virtanen}, \citenamefont {Silaev},\ and\ \citenamefont {Heikkilä}}]{lu_coupling_2022}%
  \BibitemOpen
  \bibfield  {author} {\bibinfo {author} {\bibfnamefont {Y.}~\bibnamefont {Lu}}, \bibinfo {author} {\bibfnamefont {R.}~\bibnamefont {Ojajärvi}}, \bibinfo {author} {\bibfnamefont {P.}~\bibnamefont {Virtanen}}, \bibinfo {author} {\bibfnamefont {M.~A.}\ \bibnamefont {Silaev}}, \ and\ \bibinfo {author} {\bibfnamefont {T.~T.}\ \bibnamefont {Heikkilä}},\ }\bibfield  {title} {\bibinfo {title} {Coupling the {H}iggs mode and ferromagnetic resonance in spin-split superconductors with {R}ashba spin-orbit coupling},\ }\href {\doibase 10.1103/PhysRevB.106.024514} {\bibfield  {journal} {\bibinfo  {journal} {Physical Review B}\ }\textbf {\bibinfo {volume} {106}},\ \bibinfo {pages} {024514} (\bibinfo {year} {2022}{\natexlab{b}})}\BibitemShut {NoStop}%
\bibitem [{\citenamefont {Sarma}(1963)}]{SARMA19631029}%
  \BibitemOpen
  \bibfield  {author} {\bibinfo {author} {\bibfnamefont {G.}~\bibnamefont {Sarma}},\ }\bibfield  {title} {\bibinfo {title} {On the influence of a uniform exchange field acting on the spins of the conduction electrons in a superconductor},\ }\href {\doibase https://doi.org/10.1016/0022-3697(63)90007-6} {\bibfield  {journal} {\bibinfo  {journal} {Journal of Physics and Chemistry of Solids}\ }\textbf {\bibinfo {volume} {24}},\ \bibinfo {pages} {1029} (\bibinfo {year} {1963})}\BibitemShut {NoStop}%
\bibitem [{\citenamefont {Abrikosov}(2017)}]{AbrikosovBook}%
  \BibitemOpen
  \bibfield  {author} {\bibinfo {author} {\bibfnamefont {A.}~\bibnamefont {Abrikosov}},\ }\href {https://books.google.it/books?id=tTo2DwAAQBAJ} {\emph {\bibinfo {title} {Fundamentals of the Theory of Metals}}}\ (\bibinfo  {publisher} {Dover Publications},\ \bibinfo {year} {2017})\BibitemShut {NoStop}%
\bibitem [{\citenamefont {Tewari}\ \emph {et~al.}(2011)\citenamefont {Tewari}, \citenamefont {Stanescu}, \citenamefont {Sau},\ and\ \citenamefont {Sarma}}]{Tewari_2011}%
  \BibitemOpen
  \bibfield  {author} {\bibinfo {author} {\bibfnamefont {S.}~\bibnamefont {Tewari}}, \bibinfo {author} {\bibfnamefont {T.~D.}\ \bibnamefont {Stanescu}}, \bibinfo {author} {\bibfnamefont {J.~D.}\ \bibnamefont {Sau}}, \ and\ \bibinfo {author} {\bibfnamefont {S.~D.}\ \bibnamefont {Sarma}},\ }\bibfield  {title} {\bibinfo {title} {Topologically non-trivial superconductivity in spin{\textendash}orbit-coupled systems: bulk phases and quantum phase transitions},\ }\href {\doibase 10.1088/1367-2630/13/6/065004} {\bibfield  {journal} {\bibinfo  {journal} {New Journal of Physics}\ }\textbf {\bibinfo {volume} {13}},\ \bibinfo {pages} {065004} (\bibinfo {year} {2011})}\BibitemShut {NoStop}%
\bibitem [{\citenamefont {Ketterson}\ \emph {et~al.}(1999)\citenamefont {Ketterson}, \citenamefont {Ketterson}, \citenamefont {Song},\ and\ \citenamefont {B}}]{KettersonBook}%
  \BibitemOpen
  \bibfield  {author} {\bibinfo {author} {\bibfnamefont {J.}~\bibnamefont {Ketterson}}, \bibinfo {author} {\bibfnamefont {J.}~\bibnamefont {Ketterson}}, \bibinfo {author} {\bibfnamefont {S.}~\bibnamefont {Song}}, \ and\ \bibinfo {author} {\bibfnamefont {K.}~\bibnamefont {B}},\ }\href {https://books.google.it/books?id=PdEu62yEjg0C} {\emph {\bibinfo {title} {Superconductivity}}}\ (\bibinfo  {publisher} {Cambridge University Press},\ \bibinfo {year} {1999})\BibitemShut {NoStop}%
\bibitem [{\citenamefont {Genwang~Fan}(2022)}]{Fan:52502}%
  \BibitemOpen
  \bibfield  {author} {\bibinfo {author} {\bibfnamefont {P.~Z.}\ \bibnamefont {Genwang~Fan}, \bibfnamefont {Xiao-Long~Chen}},\ }\bibfield  {title} {\bibinfo {title} {Probing two higgs oscillations in a one-dimensional fermi superfluid with raman-type spin-orbit coupling},\ }\href {\doibase 10.1007/s11467-022-1155-4} {\bibfield  {journal} {\bibinfo  {journal} {Frontiers of Physics}\ }\textbf {\bibinfo {volume} {17}},\ \bibinfo {eid} {52502} (\bibinfo {year} {2022})}\BibitemShut {NoStop}%
\bibitem [{\citenamefont {Ivakhnenko}\ \emph {et~al.}(2023)\citenamefont {Ivakhnenko}, \citenamefont {Shevchenko},\ and\ \citenamefont {Nori}}]{ivakhnenko_nonadiabatic_2023}%
  \BibitemOpen
  \bibfield  {author} {\bibinfo {author} {\bibfnamefont {O.~V.}\ \bibnamefont {Ivakhnenko}}, \bibinfo {author} {\bibfnamefont {S.~N.}\ \bibnamefont {Shevchenko}}, \ and\ \bibinfo {author} {\bibfnamefont {F.}~\bibnamefont {Nori}},\ }\bibfield  {title} {\bibinfo {title} {Nonadiabatic {L}andau-{Z}ener-{S}tückelberg-{M}ajorana transitions, dynamics, and interference},\ }\href {\doibase 10.1016/j.physrep.2022.10.002} {\bibfield  {journal} {\bibinfo  {journal} {Physics Reports}\ }\textbf {\bibinfo {volume} {995}},\ \bibinfo {pages} {1} (\bibinfo {year} {2023})}\BibitemShut {NoStop}%
\bibitem [{\citenamefont {Behrle}\ \emph {et~al.}(2018)\citenamefont {Behrle}, \citenamefont {Harrison}, \citenamefont {Kombe}, \citenamefont {Gao}, \citenamefont {Link}, \citenamefont {Bernier}, \citenamefont {Kollath},\ and\ \citenamefont {K{\"o}hl}}]{behrle2018higgs}%
  \BibitemOpen
  \bibfield  {author} {\bibinfo {author} {\bibfnamefont {A.}~\bibnamefont {Behrle}}, \bibinfo {author} {\bibfnamefont {T.}~\bibnamefont {Harrison}}, \bibinfo {author} {\bibfnamefont {J.}~\bibnamefont {Kombe}}, \bibinfo {author} {\bibfnamefont {K.}~\bibnamefont {Gao}}, \bibinfo {author} {\bibfnamefont {M.}~\bibnamefont {Link}}, \bibinfo {author} {\bibfnamefont {J.-S.}\ \bibnamefont {Bernier}}, \bibinfo {author} {\bibfnamefont {C.}~\bibnamefont {Kollath}}, \ and\ \bibinfo {author} {\bibfnamefont {M.}~\bibnamefont {K{\"o}hl}},\ }\bibfield  {title} {\bibinfo {title} {Higgs mode in a strongly interacting fermionic superfluid},\ }\href {https://doi.org/10.1038/s41567-018-0128-6} {\bibfield  {journal} {\bibinfo  {journal} {Nature Physics}\ }\textbf {\bibinfo {volume} {14}},\ \bibinfo {pages} {781} (\bibinfo {year} {2018})}\BibitemShut {NoStop}%
\bibitem [{\citenamefont {Wang}\ \emph {et~al.}(2015)\citenamefont {Wang}, \citenamefont {Yi},\ and\ \citenamefont {Xianlong}}]{Wang_2015}%
  \BibitemOpen
  \bibfield  {author} {\bibinfo {author} {\bibfnamefont {P.}~\bibnamefont {Wang}}, \bibinfo {author} {\bibfnamefont {W.}~\bibnamefont {Yi}}, \ and\ \bibinfo {author} {\bibfnamefont {G.}~\bibnamefont {Xianlong}},\ }\bibfield  {title} {\bibinfo {title} {Topological phase transition in the quench dynamics of a one-dimensional fermi gas with spin-orbit coupling},\ }\href {\doibase 10.1088/1367-2630/17/1/013029} {\bibfield  {journal} {\bibinfo  {journal} {New Journal of Physics}\ }\textbf {\bibinfo {volume} {17}},\ \bibinfo {pages} {013029} (\bibinfo {year} {2015})}\BibitemShut {NoStop}%
\bibitem [{\citenamefont {Dong}\ \emph {et~al.}(2015)\citenamefont {Dong}, \citenamefont {Dong}, \citenamefont {Gong},\ and\ \citenamefont {Pu}}]{dong2015dynamical}%
  \BibitemOpen
  \bibfield  {author} {\bibinfo {author} {\bibfnamefont {Y.}~\bibnamefont {Dong}}, \bibinfo {author} {\bibfnamefont {L.}~\bibnamefont {Dong}}, \bibinfo {author} {\bibfnamefont {M.}~\bibnamefont {Gong}}, \ and\ \bibinfo {author} {\bibfnamefont {H.}~\bibnamefont {Pu}},\ }\bibfield  {title} {\bibinfo {title} {Dynamical phases in quenched spin--orbit-coupled degenerate fermi gas},\ }\href@noop {} {\bibfield  {journal} {\bibinfo  {journal} {Nature communications}\ }\textbf {\bibinfo {volume} {6}},\ \bibinfo {pages} {6103} (\bibinfo {year} {2015})}\BibitemShut {NoStop}%
\bibitem [{\citenamefont {Gor'kov}\ and\ \citenamefont {Rashba}(2001)}]{gorkov_superconducting_2001}%
  \BibitemOpen
  \bibfield  {author} {\bibinfo {author} {\bibfnamefont {L.~P.}\ \bibnamefont {Gor'kov}}\ and\ \bibinfo {author} {\bibfnamefont {E.~I.}\ \bibnamefont {Rashba}},\ }\bibfield  {title} {\bibinfo {title} {Superconducting 2d system with lifted spin degeneracy: Mixed singlet-triplet state},\ }\href {\doibase 10.1103/PhysRevLett.87.037004} {\bibfield  {journal} {\bibinfo  {journal} {Physical Review Letters}\ }\textbf {\bibinfo {volume} {87}},\ \bibinfo {pages} {037004} (\bibinfo {year} {2001})}\BibitemShut {NoStop}%
\bibitem [{\citenamefont {Frigeri}\ \emph {et~al.}(2004)\citenamefont {Frigeri}, \citenamefont {Agterberg},\ and\ \citenamefont {Sigrist}}]{Frigeri_2004}%
  \BibitemOpen
  \bibfield  {author} {\bibinfo {author} {\bibfnamefont {P.~A.}\ \bibnamefont {Frigeri}}, \bibinfo {author} {\bibfnamefont {D.~F.}\ \bibnamefont {Agterberg}}, \ and\ \bibinfo {author} {\bibfnamefont {M.}~\bibnamefont {Sigrist}},\ }\bibfield  {title} {\bibinfo {title} {Spin susceptibility in superconductors without inversion symmetry},\ }\href {\doibase 10.1088/1367-2630/6/1/115} {\bibfield  {journal} {\bibinfo  {journal} {New Journal of Physics}\ }\textbf {\bibinfo {volume} {6}},\ \bibinfo {pages} {115} (\bibinfo {year} {2004})}\BibitemShut {NoStop}%
\bibitem [{\citenamefont {Szombati}\ \emph {et~al.}(2016)\citenamefont {Szombati}, \citenamefont {Nadj-Perge}, \citenamefont {Car}, \citenamefont {Plissard}, \citenamefont {Bakkers},\ and\ \citenamefont {Kouwenhoven}}]{szombati_josephson_2016}%
  \BibitemOpen
  \bibfield  {author} {\bibinfo {author} {\bibfnamefont {D.~B.}\ \bibnamefont {Szombati}}, \bibinfo {author} {\bibfnamefont {S.}~\bibnamefont {Nadj-Perge}}, \bibinfo {author} {\bibfnamefont {D.}~\bibnamefont {Car}}, \bibinfo {author} {\bibfnamefont {S.~R.}\ \bibnamefont {Plissard}}, \bibinfo {author} {\bibfnamefont {E.~P. A.~M.}\ \bibnamefont {Bakkers}}, \ and\ \bibinfo {author} {\bibfnamefont {L.~P.}\ \bibnamefont {Kouwenhoven}},\ }\bibfield  {title} {\bibinfo {title} {Josephson $\phi$-junction in nanowire quantum dots},\ }\href {\doibase 10.1038/nphys3742} {\bibfield  {journal} {\bibinfo  {journal} {Nature Physics}\ }\textbf {\bibinfo {volume} {12}},\ \bibinfo {pages} {568} (\bibinfo {year} {2016})}\BibitemShut {NoStop}%
\bibitem [{\citenamefont {Kirilyuk}\ \emph {et~al.}(2010)\citenamefont {Kirilyuk}, \citenamefont {Kimel},\ and\ \citenamefont {Rasing}}]{kirilyuk_ultrafast_2010}%
  \BibitemOpen
  \bibfield  {author} {\bibinfo {author} {\bibfnamefont {A.}~\bibnamefont {Kirilyuk}}, \bibinfo {author} {\bibfnamefont {A.~V.}\ \bibnamefont {Kimel}}, \ and\ \bibinfo {author} {\bibfnamefont {T.}~\bibnamefont {Rasing}},\ }\bibfield  {title} {\bibinfo {title} {Ultrafast optical manipulation of magnetic order},\ }\href {\doibase 10.1103/RevModPhys.82.2731} {\bibfield  {journal} {\bibinfo  {journal} {Reviews of Modern Physics}\ }\textbf {\bibinfo {volume} {82}},\ \bibinfo {pages} {2731} (\bibinfo {year} {2010})}\BibitemShut {NoStop}%
\bibitem [{\citenamefont {Kirilyuk}\ \emph {et~al.}(2013)\citenamefont {Kirilyuk}, \citenamefont {Kimel},\ and\ \citenamefont {Rasing}}]{kirilyuk_laser-induced_2013}%
  \BibitemOpen
  \bibfield  {author} {\bibinfo {author} {\bibfnamefont {A.}~\bibnamefont {Kirilyuk}}, \bibinfo {author} {\bibfnamefont {A.~V.}\ \bibnamefont {Kimel}}, \ and\ \bibinfo {author} {\bibfnamefont {T.}~\bibnamefont {Rasing}},\ }\bibfield  {title} {\bibinfo {title} {Laser-induced magnetization dynamics and reversal in ferrimagnetic alloys},\ }\href {\doibase 10.1088/0034-4885/76/2/026501} {\bibfield  {journal} {\bibinfo  {journal} {Reports on Progress in Physics}\ }\textbf {\bibinfo {volume} {76}},\ \bibinfo {pages} {026501} (\bibinfo {year} {2013})}\BibitemShut {NoStop}%
\bibitem [{\citenamefont {El-Ghazaly}\ \emph {et~al.}(2020)\citenamefont {El-Ghazaly}, \citenamefont {Gorchon}, \citenamefont {Wilson}, \citenamefont {Pattabi},\ and\ \citenamefont {Bokor}}]{ELGHAZALY2020166478}%
  \BibitemOpen
  \bibfield  {author} {\bibinfo {author} {\bibfnamefont {A.}~\bibnamefont {El-Ghazaly}}, \bibinfo {author} {\bibfnamefont {J.}~\bibnamefont {Gorchon}}, \bibinfo {author} {\bibfnamefont {R.~B.}\ \bibnamefont {Wilson}}, \bibinfo {author} {\bibfnamefont {A.}~\bibnamefont {Pattabi}}, \ and\ \bibinfo {author} {\bibfnamefont {J.}~\bibnamefont {Bokor}},\ }\bibfield  {title} {\bibinfo {title} {Progress towards ultrafast spintronics applications},\ }\href {\doibase https://doi.org/10.1016/j.jmmm.2020.166478} {\bibfield  {journal} {\bibinfo  {journal} {Journal of Magnetism and Magnetic Materials}\ }\textbf {\bibinfo {volume} {502}},\ \bibinfo {pages} {166478} (\bibinfo {year} {2020})}\BibitemShut {NoStop}%
\bibitem [{\citenamefont {Plekhanov}\ \emph {et~al.}(2019)\citenamefont {Plekhanov}, \citenamefont {Thakurathi}, \citenamefont {Loss},\ and\ \citenamefont {Klinovaja}}]{plekhanov_floquet_2019}%
  \BibitemOpen
  \bibfield  {author} {\bibinfo {author} {\bibfnamefont {K.}~\bibnamefont {Plekhanov}}, \bibinfo {author} {\bibfnamefont {M.}~\bibnamefont {Thakurathi}}, \bibinfo {author} {\bibfnamefont {D.}~\bibnamefont {Loss}}, \ and\ \bibinfo {author} {\bibfnamefont {J.}~\bibnamefont {Klinovaja}},\ }\bibfield  {title} {\bibinfo {title} {Floquet second-order topological superconductor driven via ferromagnetic resonance},\ }\href {\doibase 10.1103/PhysRevResearch.1.032013} {\bibfield  {journal} {\bibinfo  {journal} {Physical Review Research}\ }\textbf {\bibinfo {volume} {1}},\ \bibinfo {pages} {032013} (\bibinfo {year} {2019})}\BibitemShut {NoStop}%
\bibitem [{\citenamefont {Gershenzon}\ \emph {et~al.}(1990)\citenamefont {Gershenzon}, \citenamefont {Gershenzon}, \citenamefont {Gol'tsman}, \citenamefont {Lyul'kin}, \citenamefont {Semenov},\ and\ \citenamefont {Sergeev}}]{gershenzon1990electron}%
  \BibitemOpen
  \bibfield  {author} {\bibinfo {author} {\bibfnamefont {E.}~\bibnamefont {Gershenzon}}, \bibinfo {author} {\bibfnamefont {M.}~\bibnamefont {Gershenzon}}, \bibinfo {author} {\bibfnamefont {G.}~\bibnamefont {Gol'tsman}}, \bibinfo {author} {\bibfnamefont {A.}~\bibnamefont {Lyul'kin}}, \bibinfo {author} {\bibfnamefont {A.}~\bibnamefont {Semenov}}, \ and\ \bibinfo {author} {\bibfnamefont {A.}~\bibnamefont {Sergeev}},\ }\bibfield  {title} {\bibinfo {title} {Electron-phonon interaction in ultrathin $\text{Nb}$ films},\ }\href@noop {} {\bibfield  {journal} {\bibinfo  {journal} {Sov. Phys. JETP}\ }\textbf {\bibinfo {volume} {70}},\ \bibinfo {pages} {505} (\bibinfo {year} {1990})}\BibitemShut {NoStop}%
\bibitem [{\citenamefont {Kardakova}\ \emph {et~al.}(2013)\citenamefont {Kardakova}, \citenamefont {Finkel}, \citenamefont {Morozov}, \citenamefont {Kovalyuk}, \citenamefont {An}, \citenamefont {Dunscombe}, \citenamefont {Tarkhov}, \citenamefont {Mauskopf}, \citenamefont {Klapwijk},\ and\ \citenamefont {Goltsman}}]{kardakova_electron-phonon_2013}%
  \BibitemOpen
  \bibfield  {author} {\bibinfo {author} {\bibfnamefont {A.}~\bibnamefont {Kardakova}}, \bibinfo {author} {\bibfnamefont {M.}~\bibnamefont {Finkel}}, \bibinfo {author} {\bibfnamefont {D.}~\bibnamefont {Morozov}}, \bibinfo {author} {\bibfnamefont {V.}~\bibnamefont {Kovalyuk}}, \bibinfo {author} {\bibfnamefont {P.}~\bibnamefont {An}}, \bibinfo {author} {\bibfnamefont {C.}~\bibnamefont {Dunscombe}}, \bibinfo {author} {\bibfnamefont {M.}~\bibnamefont {Tarkhov}}, \bibinfo {author} {\bibfnamefont {P.}~\bibnamefont {Mauskopf}}, \bibinfo {author} {\bibfnamefont {T.~M.}\ \bibnamefont {Klapwijk}}, \ and\ \bibinfo {author} {\bibfnamefont {G.}~\bibnamefont {Goltsman}},\ }\bibfield  {title} {\bibinfo {title} {The electron-phonon relaxation time in thin superconducting titanium nitride films},\ }\href {\doibase 10.1063/1.4851235} {\bibfield  {journal} {\bibinfo  {journal} {Applied Physics Letters}\ }\textbf {\bibinfo {volume} {103}},\ \bibinfo {pages} {252602} (\bibinfo {year} {2013})}\BibitemShut {NoStop}%
\bibitem [{\citenamefont {Tsuchiya}\ \emph {et~al.}(2018)\citenamefont {Tsuchiya}, \citenamefont {Yamamoto}, \citenamefont {Yoshii},\ and\ \citenamefont {Nitta}}]{tsuchiya_hidden_2018}%
  \BibitemOpen
  \bibfield  {author} {\bibinfo {author} {\bibfnamefont {S.}~\bibnamefont {Tsuchiya}}, \bibinfo {author} {\bibfnamefont {D.}~\bibnamefont {Yamamoto}}, \bibinfo {author} {\bibfnamefont {R.}~\bibnamefont {Yoshii}}, \ and\ \bibinfo {author} {\bibfnamefont {M.}~\bibnamefont {Nitta}},\ }\bibfield  {title} {\bibinfo {title} {Hidden charge-conjugation, parity, and time-reversal symmetries and massive {G}oldstone ({H}iggs) modes in superconductors},\ }\href {\doibase 10.1103/PhysRevB.98.094503} {\bibfield  {journal} {\bibinfo  {journal} {Physical Review B}\ }\textbf {\bibinfo {volume} {98}},\ \bibinfo {pages} {094503} (\bibinfo {year} {2018})}\BibitemShut {NoStop}%
\end{thebibliography}
\end{document}